\documentstyle{l-aa}
 \input psfig.sty 
 \def \hi {H\,{\sc i~}} 
 \begin{document} 
 \thesaurus{04(09.03.1, 10.08.1, 10.11.1, 13.25.4)} 
 \title{A search for soft X-ray emission associated with prominent 
high-velocity-cloud complexes} 
 \author{J. Kerp\inst{1,2}, W. B. Burton\inst{3}, R. Egger\inst{1}, M.J. 
 Freyberg\inst{1}, Dap Hartmann\inst{4,2}, P.M.W. Kalberla\inst{2}, U. 
 Mebold\inst{2}, J. Pietz\inst{2} } 
 
 \institute{ Max-Planck-Institut f\"ur Extraterrestrische Physik, Postfach 
 1603, D-85740 Garching, Germany \and Radioastronomisches Institut der 
 Universit\"at Bonn, Auf dem H\"ugel 71, D-53121 Bonn, Germany \and 
 Sterrewacht Leiden, P.O. Box 9513, NL-2300 RA Leiden, The Netherlands \and 
 Harvard-Smithsonian Center for Astrophysics, 60 Garden Street, Cambridge, 
 MA 02138, U.S.A.} 
 
 \date{Received xxxx / xxxx 1998} 
 \offprints{J. Kerp, Bonn address} 
 
 \maketitle \markboth{J. Kerp et al.}{The SXRB towards HVC complexes} 
 
 \begin{abstract} {We correlate the {\it ROSAT\/} 
 $\frac{1}{4}$\,keV all-sky survey with the Leiden/Dwingeloo 
 \hi survey, looking for soft X-ray signatures of prominent 
 high-velocity-cloud (HVC) complexes.  We study the transfer of 
 $\frac{1}{4}$\,keV photons through the interstellar medium
 in order to distinguish variations in the soft X-ray background (SXRB) 
 intensity caused by photoelectric absorption effects from those due to 
 excess X-ray emission. The X-ray data are modelled as a combination of 
emission from the Local Hot Bubble (LHB) and emission from a distant plasma 
in the galactic halo and extragalactic sources. The X-ray radiation intensity
of the galactic halo and extragalactic X-ray background is modulated by 
the photoelectric absorption of the intervening galactic interstellar matter.
We show that large- and small-scale intensity variations of the 
$\frac{1}{4}$\,keV SXRB are caused by photoelectric absorption which is 
predominantly traced by the total $N_{\rm HI}$ distribution. The extensive 
coverage of the two surveys supports evidence for a hot, X-ray emitting 
corona. We show that this leads to a good representation of the SXRB 
observations.  For four large areas on the sky, we search for regions where 
the modelled and observed X-ray emission differ.  We find that there is 
excess X-ray emission towards regions near HVC complexes C, D, and GCN.
We suggest that the excess X-ray emission 
is positionally correlated with the high-velocity clouds.  Some lines of 
sight towards HVCs also pass through significant amounts of 
intermediate-velocity gas, so we cannot constrain the possible role played 
by IVC gas in these directions of HVC and IVC overlap, in determining the 
X-ray excesses.  } 
 
 \keywords{ISM: clouds -- Galaxy: halo -- Galaxy: 
 kinematics and dynamics -- X-rays: ISM} \end{abstract} 
 
 \section{Introduction} High-velocity clouds are \hi structures 
 characterized by radial velocities which deviate typically by several 
hundred km s$^{-1}$ from conventional galactic 
 rotation (see Wakker \& van Woerden (1997) for a recent review). 
 Distances remain uncertain for most of the clouds.  Distance 
 limits have been constrained for only two lines of sight by optical 
 absorption lines found by Danly et al. (1993) and by van Woerden et al. 
 (1998) toward complexes M ($d < 5$\,kpc) and A ($4 < d < 10$\,kpc). There is consensus that the HVCs comprising the Magellanic 
 Stream are at Magellanic Cloud distances ($\simeq$\,50\,kpc), based on
 positional 
 and kinematic coincidences and the ability of tidal models to account for 
these coincidences.  But the matter of distances remains largely 
 unresolved for the majority of the HVCs.  Blitz et al. (1998) suggest that 
 some HVCs are scattered throughout the Local Group, excepting the principal 
 northern complexes and the Magellanic Stream.  Morphological arguments have 
 led several authors (e.g. Hirth et al. 1985) to suggest that some HVCs 
 interact with the galactic disk. 
 This scenario is supported by the detection of soft X-ray enhancements 
 close to HVC complexes M (Herbstmeier et al. 1995) and C (Hirth et al. 
 1985; Kerp et al. 1994, 1995, 1996). In addition, evidence of a physical 
 connection of some HVCs with the galactic disk has 
 been found in the \hi ``velocity bridges'' which seem to link the HVC 
 gas with gas at conventional velocities (Pietz et al. 1996). 
 
 We extend the SXRB investigations of Herbstmeier et al. (1995) and Kerp et 
 al. (1996) to other HVC complexes, different in location, velocity, and 
 possibly in origin, we use correlations of {\it ROSAT\/} 
 $\frac{1}{4}$\,keV X-ray data (see Snowden et al. 1997) with data from the 
 Leiden/Dwingeloo \hi survey (Hartmann \& Burton 1997). 
 The selected fields are at high latitudes, widely distributed over the sky, 
 which encompass readily identifiable (see Wakker \& van Woerden 1997) parts 
 of HVC complexes. The complexes C, A, D, WA, and GCN
 fit those criteria; the detailed shapes of 
 the selected fields were partly determined by the polar-grid projection of 
 the {\it ROSAT\/} data. 
 
 We evaluate the transmission of $\frac{1}{4}$\,keV photons through the 
 X-ray absorbing interstellar medium, and demonstrate that the transmission 
 is quantitatively traced by \hi. Our approach aims at distinguishing  fluctuations in the soft X-ray intensities caused by 
 photoelectric absorption effects from those signifying true excess soft 
 X-ray emission. To this end, we first model the SXRB distribution modulated 
 by the photoelectric absorption caused by Milky Way gas at 
 conventional and intermediate velocities.  We then subtract the modelled 
 SXRB distribution from the observed one, and identify regions 
 where the modelled distribution deviates from what is observed.  
 
 In Section 2, we describe the 
 X-ray and \hi data used. In Section 3, we evaluate the soft X-ray 
 radiation-transfer equation with the goal of finding HVC signatures 
 in the SXRB distribution. In Section 4, we show the results of the 
 correlation analysis towards individual HVC complexes. In Section 5, we 
 discuss the implications for the origin and distribution of the SXRB 
 sources. The results are summarized in Section 6.
 
 \section{X-ray and \hi data}
 \begin{table*}
 \caption[]{Location of the HVC fields selected, and the $N_{\rm HI}$ and 
 X-ray count rate ranges encountered in each field. The mean {\it ROSAT\/} 
 integration times, $t_{\rm X-ray}$, are also given, with minimum and maximum 
 times noted in parentheses.}
 \begin{flushleft}
 \begin{tabular}{lcclll}
 \hline
 complex & {\it l}--range   & {\it b}--range & $N_{\rm HI}$ & $I_{\rm 1/4\,keV}$ & $t_{\rm X-ray}$\\
 & & & ($10^{20} {\rm cm}^{-2}$) & ($10^{-4} {\rm cts\, s^{-1} 
 arcmin^{-2}}$) & (seconds) \\
 \hline
 GCN         & 18$\degr$ -- 73$\degr$     & $-$52$\degr$ -- $-$15$\degr$ & 2.0 -- 12.0 & 0.7 -- 12.0 & 460~(170\,--\,680)\\
 C low, D     & 34$\degr$ -- 86$\degr$     & +33$\degr$  -- +79$\degr$ & 0.6 -- 11.5 & 2.6 -- 20.8 & 1000~(262\,--\,3200)\\
 WA          & 218$\degr$ -- 270$\degr$   & +24$\degr$  -- +52$\degr$ &  1.5 -- 7.9 & 3.0 -- 12.7 & 530~(331\,--\,714)\\
 C high      & 99$\degr$ -- 166$\degr$    & +12$\degr$ --  +74$\degr$ &  
0.3 -- 19.3 & 2.1 -- 44.6 & 750~(244\,--\,3514)\\
 \hline
 \end{tabular}
 \end{flushleft}
 
 \end{table*}

 The X-ray data were obtained from the {\it ROSAT\/} all-sky survey (Snowden 
 \& Schmitt 1990; Voges 1992; Snowden et al. 1997). Photon events detected 
 by the Position Sensitive Proportional Counter (PSPC: Pfeffermann et al. 
 1986) were binned into seven pulse-height channels (R1 -- R7: Snowden et 
 al. 1994a) covering the entire {\it ROSAT\/} PSPC energy window. The SXRB 
 radiation between $0.1\,{\rm keV}\,\leq\,E\,\leq\,0.28$\,keV was measured 
 in the R1 and R2 bands. Combining the R1 and R2 bands to produce the 
 {\it ROSAT\/} $\frac{1}{4}$\,keV data offers the highest statistical 
 significance of soft X-ray material available. The $\frac{1}{4}$\,keV 
 energy range is the most sensitive of the {\it ROSAT\/} PSPC bands to 
 photoelectric absorption by the interstellar medium. In this band the 
 interstellar absorption cross section is about $\sigma_{\rm 
 X}\,\simeq\,10^{-20}\,{\rm cm^2\,\hi^{-1}}$. In consequence, the product of 
 soft X-ray absorption cross section and the \hi column density, $N_{\rm 
 HI}$, is close to or greater than unity across the sky, with the exception 
 of a few lines of sight. The data are corrected for scattered solar X-rays 
 (Snowden \& Freyberg 1993), as well as for particle background (Plucinsky 
 et al. 1993) and long-term X-ray enhancements (Snowden et al. 1995).
 The full intrinsic angular resolution of the PSPC has been used, yielding 
 maps with $12\arcmin$ resolution; point sources have been removed to a 
 minimum count rate of $0.02\,\mbox{cts}\,\mbox{s}^{-1}$ (Snowden et al. 
 1997).
 
 The \hi data are those of the Leiden/Dwingeloo survey of Hartmann \& Burton 
 (1997), who used the Dwingeloo 25-m telescope to observe the sky at $\delta 
 \geq -30\degr$ with a true-angle grid spacing of $0\fdg5$ in both $l$ and 
 $b$. The velocity resolution is set by the interval of 1.03 km s$^{-1}$ 
 between each of the 1024 channels of the spectrometer; the material covers 
 LSR velocities between $-450$ km s$^{-1}$ and $+400$ km s$^{-1}$, and thus 
 encompasses essentially all HVC emission. The rms limit on the measured 
 brightness-temperature fluctuations is $\Delta T_{\rm B}\,=\,0.07$\,K. The 
 correction for stray radiation is described by 
 Hartmann et al. (1996). The \hi data are published as FITS files on 
 a CD-ROM by Hartmann \& Burton (1997), together with an atlas of maps.  
 
 Table 1 summarizes the main parameters of the regions studied as well as 
 their typical X-ray intensities and \hi column densities. 
 We projected the $N_{\rm HI}$ distribution, regridded to an angular 
 resolution of $48 \arcmin$, onto the polar-grid projection of the {\it 
 ROSAT\/} survey. The choice of angular resolution aimed at enhancing the 
 statistical significance of the X-ray data and allowing differentiation 
 between systematic uncertainties introduced by X-ray raw-data processing 
 (e.g. residual point source contributions and scanning stripes) and 
 modelling of the X-ray intensity distribution. The statistical 
 significance, $\sigma$ (corresponding to the uncertainty within a 
 $48\arcmin\,\times\,48\arcmin$ area), of soft X-ray enhancements and 
 depressions was evaluated using the {\it ROSAT\/} uncertainty maps, which 
 account only for the number of photon events: they do not include any 
 systematic uncertainties introduced by non-cosmic 
 X-ray backgrounds. 
 
 \section{Radiation transfer of the soft X-rays}
 
 Earlier investigations of the {\it ROSAT\/} data (Snowden
 et al. 1994b; Herbstmeier et al. 1995; Kerp et al. 1996) indicated that the 
 brightnesses from both, the distant X-ray
 sources and from the Local Hot Bubble (LHB) vary across the sky. Because of 
 the variations of both source terms, we first address some general 
 properties of the $\frac{1}{4}$\,keV radiation transfer through the 
 interstellar medium before attempting to identify imprints of HVCs on the 
 SXRB radiation.
 We focus on the following questions: 
 
 \begin{enumerate}
 \item Are there significant variations of the SXRB source distribution on 
 scales of $0\fdg8$ to tens of degrees?
 \item Is \hi alone the tracer of the soft X-ray absorption by neutral 
 matter? Do the \hi/X-ray results require diffusely distributed ${\rm H_2}$ 
 and/or ${\rm H^+}$ at high $|b|$ as additional tracers of soft X-ray 
 absorption? 
 \item Can the velocity range of galactic \hi accounting for the SXRB 
 absorption be constrained: does the gas at conventional and intermediate 
 galactic velocities suffice, or are HVCs also implicated?
 \end{enumerate}
 To answer these questions, we sought an expression for the soft X-ray 
 radiation-transfer equation which reveals simultaneously the intensity and 
 positional distribution of the individual source terms.
 
 \subsection{The soft X-ray radiation-transfer equation} 
 
 The soft X-ray intensity distribution is modulated by photoelectric 
 absorption due to interstellar matter lying between the observer and the 
 source of the X-rays. The effective photoelectric absorption cross section 
 ($\sigma_{\rm X}$) depends on the chemical composition of the absorbing 
 matter, normalized to a mean absorption cross section per neutral hydrogen 
 atom (Morrison \& McCammon 1983). Moreover, the absolute value of this 
 cross section depends on the source spectrum and on the sensitivity 
 function (bandpass) of the X-ray detector system. These dependences stem 
 from the energy dependence of the absorption cross section ($\sigma_{\rm 
 X}\,\propto\,E^{-\frac{8}{3}}$). This leads to stronger attenuation for the 
 lower-energy X-ray photons than for the more energetic ones. The more 
 absorbing matter is located along the line of sight the stronger the 
 softer-energy end is attenuated relative to the harder-energy end.  This 
 situation leads to an apparent hardening of the source X-ray spectrum due 
 to photoelectric absorption. 
 
 The dependence of photoelectric absorption cross section on $N_{\rm HI}$ is 
 shown in Fig.\,\ref{fig:cross} for the LHB, for a galactic halo plasma, and 
 for a power-law extragalactic X-ray spectrum.  We discuss below the X-ray 
 source spectra of these three components.  At high $|b|$, the {\it ROSAT\/} 
 PSPC data suggest that, in addition to emission from the LHB, diffusely 
 distributed X-ray emission originates beyond the bulk of the galactic \hi 
 gas layer (Herbstmeier et al. 1995; Kerp et al. 1996; Pietz et al. 1998a, 
 1998b; Wang 1998). This situation requires at least two source terms in the 
 radiation-transfer equation.  
 
 The X-ray emission from the LHB evidently originates from a thermal plasma 
 ($I_{\rm LHB}$; McCammon \& Sanders 1990),
 embedded in the local void of neutral matter.  The local interstellar 
 cavity is evidently an irregularly-shaped, low-volume-density region 
 enclosing the solar neighborhood, where the X-ray intensity varies 
 (roughly proportionally to the pathlength, in the range of 50 to 150 pc, 
 through the local cavity) on scales of several tens of degrees (Cox \& 
 Reynolds 1987; Egger et al. 1996). 
 
 The distant soft X-ray emission is most likely the superposition of thermal 
 plasma radiation ($I_{\rm halo}$; Kerp 1994; Sidher et al. 1996) from the 
 galactic halo (Pietz et al. 1998a, 1998b) and emission from unresolved 
 extragalactic point sources building up the extragalactic soft X-ray 
 background ($I_{\rm extragal}$; Hasinger et al. 1993).
 Accordingly, the soft X-ray transfer equation has the form:
 \begin{eqnarray}
 I_{\rm SXRB}\,&=&\,I_{\rm LHB}\cdot e^{-\sigma_{\rm X}{\rm (LHB)}\cdot
 N_{\rm HI}{\rm (LHB)}}\,+\nonumber \\ 
 & & +\,I_{\rm halo}\cdot e^{-\sigma_{\rm X}{\rm (halo)}\cdot N_{\rm
 HI}({\rm total})}\,+ \\
 & & \mbox{} +\,I_{\rm extragal}\cdot e^{-\sigma_{\rm X}{\rm 
 (extragal)}\cdot N_{\rm HI}{\rm (total)}\nonumber}\end{eqnarray}
 
 \subsection{The spectral properties of the distant source terms}
 
 The LHB term represents the thermal plasma radiation of the local gas. The 
 intensity of the LHB varies across the entire sky: $I_{\rm 
 LHB}$\,=\,(2.5\,--\,8.2)$\cdot10^{-4}\,{\rm cts\,s^{-1} arcmin^{-2}}$ 
 (Snowden et al. 1998). The distant soft X-ray term represents the 
 superposition of the isotropically distributed intensity of the 
 extragalactic background radiation and the distant galactic plasma 
 radiation. The unabsorbed X-ray intensity contributed by the extragalactic 
 radiation is about $I_{\rm extragal}$\,=\,(2.3\,--\,4.4)$\cdot 10^{-4}\,{\rm 
 cts\,s^{-1}arcmin^{-2}}$ (Barber et al. 1996; Cui et al. 1996) while the 
 unabsorbed distant X-ray intensity (assuming a patchy galactic X-ray halo) 
 is about $I_{\rm halo}$\,=\,(4.0\,--\,30)$\cdot 10^{-4}\,{\rm 
 cts\,s^{-1}arcmin^{-2}}$ (Snowden et al. 1998). Thus, the distant galactic 
 X-ray plasma will be of prime importance in studying $I_{\rm distant}\,=\,I_{\rm halo}\,+\,I_{\rm extragal}$ 
 because it is the source term with the largest intensity range and with an 
 unknown distribution across the fields of interest.
 
 It is plausible to assume that all of the galactic ISM is available to 
 absorb radiation from the extragalactic SXRB component. Pietz et al. 
 (1998b) derive an exponential scale height of $h_z \sim 4.4$ kpc for 
 the X-ray emitting halo, while Lockman \& Gehman (1991) showed most of the 
 conventional-velocity galactic \hi gas is located at $|z| \la 0.4$\ kpc. 

 \subsubsection{The source spectrum of $I_{\rm halo}$}
 
 The galactic halo X-ray emission is evidently due to thermal-plasma 
 processes. Rocchia et al. (1984) found plasma
 emission from O$^{+6}$ and O$^{+7}$ ions. Hasinger (1991)
 found indications in deep PSPC observations for an
 emission bump in the X-ray spectrum near 0.6\,keV, also indicating the
 presence of these ions. Kerp (1994) and Sidher et al. (1996)
 showed that a thermal-plasma spectrum fits PSPC data well. The 
 PSPC data suggest that the distant X-ray plasma, approximated by 
 the Raymond \& Smith (1977) model, has a temperature $T_{\rm 
 plasma}\,=\,10^{6.3\,\pm\,0.1}$\,K. In view of these results and of those  of Pietz 
 et al. (1998b), we assume that the galactic halo plasma is in collisionally 
 ionized equilibrium. (This assumption is a simplification of the plasma 
 processes occurring at high $|z|$, but is reasonable despite lacking detailed 
 information about the X-ray spectrum.) Note that near $T \sim 10^{6.3}$\,K, 
 the absorption cross-section in the 
 $\frac{1}{4}$\,keV band does not depend strongly on plasma temperature 
 (Snowden et al. 1997).

 \subsubsection{The source spectrum of $I_{\rm extragal}$}
 
 $I_{\rm extragal}$ is caused by the superposition of X-rays from 
 extragalactic point sources (Hasinger et al. 1998). The spectrum of the 
 extragalactic background is a matter of discussion (Gendreau et al. 1995; 
 Georgantopoulos et al. 1996).  The averaged spectrum of bright, discrete 
 soft X-ray sources, together providing the extragalactic background in the 
 {\it ROSAT\/} energy window, {can be approximated by a power law ($E^{- 
 \Gamma}$) with an averaged spectral index of 2.1\,--\,2.2 (Hasinger et al. 
 1993, 1998).} At lower fluxes, the contribution of faint emission-line 
 galaxies dominates the spectral properties of the extragalactic background, 
 leading to a flatter power-law slope (Almaini et al. 1996). Our 
 investigation of the SXRB deals with 
 lower source fluxes than those investigated by Almaini et al.; accordingly, 
 a plausible value of the extragalactic spectral index is 
 $\Gamma\,\simeq\,1.5$ (Gendreau et al. 1995).
 \subsection{The simplified soft X-ray radiation-transfer equation}
 
 We show that we may simplify Eq.\,(1) into an expression involving only two 
 X-ray source terms for the $\frac{1}{4}$\,keV band, namely the LHB source 
 term ($I_{\rm LHB}$) and the distant source term ($I_{\rm distant}$), 
 representing the superposition of the thermal plasma emission beyond the 
 bulk of the galactic \hi and the extragalactic background radiation: 
 ($I_{\rm distant}\,=\,I_{\rm halo}\,+\,I_{\rm extragal}$).
 
 \subsubsection{The $I_{\rm LHB}$ source term}
 
 The LHB source term varies approximately in proportion to the extent of the 
 local cavity (Snowden et al. 1998). The {\it ROSAT\/} X-ray data considered 
 here are limited in sensitivity (at the 3-$\sigma$ level) to $N_{\rm HI}$ 
 variations of about $N_{\rm HI} \sim 5\cdot10^{19}\,{\rm cm^{-2}}$. The 
 moderate angular resolution of the data we chose limits the angular extent 
 of the small-scale intensity variations, anti-correlated to small-scale 
 $N_{\rm HI}$ variations, to about $48\arcmin$.
 Thus, a narrow \hi filament with $N_{\rm HI}\,\leq\,5\cdot 10^{19}\,{\rm 
 cm^{-2}}$ will not be detectable.
 Taking these limitations into account, interstellar absorption-line 
 measurements (Welsh 1998) show that properties of the local cavity vary 
 smoothly on angular scales of several tens of degrees. Therefore, $I_{\rm 
 LHB}$ reveals a distribution of soft X-rays approximately smooth over tens 
 of degrees. We start our analysis using the assumption that across each 
 individual field $I_{\rm LHB}$\,=\,const., and then show below that this 
 conforms to the observed situation.
 
 Because the effective photoelectric absorption cross section of the LHB 
 plasma is larger than that of the galactic-halo plasma and of 
 the extragalactic power-law spectrum (Fig.\,\ref{fig:cross}), deviations 
 from the assumption of $I_{\rm LHB}$\,=\,const. will be easily detected. A 
 local cloud attenuating soft X-rays will be disclosed by a 
 deeper soft X-ray shadow than would be the case if the same cloud were 
 located outside the LHB (see Kerp \& Pietz 1998).
 
 \subsubsection{The $I_{\rm distant}$ source term}
 
 The $I_{\rm distant}$ source term represents the sum of $I_{\rm halo}$ and 
 $I_{\rm extragal}$. Fig.\,\ref{fig:cross} suggests that the photoelectric 
 absorption cross sections of the halo plasma and the extragalactic 
 power-law spectrum have comparable values. The largest difference between 
 the cross sections, amounting to some 20\%, occurs in the range $N_{\rm 
 HI}\,=\,1\,-\,3\cdot 10^{20}\,{\rm cm^{-2}}$. Evaluating ($e^{-\sigma_{\rm 
 X}{\rm (halo,extragal)}\cdot N_{\rm HI}}$), we see that such a difference 
 corresponds to a 7\% effect on $I_{\rm distant}$, which is negligible to 
 our purposes in view of the statistical limitations of the X-ray data.
 Moreover, as we show below (see Fig.\,\ref{fig:sxrb_long}), $I_{\rm 
 halo} > I_{\rm extragal}$ in most regions of the high-$|b|$ sky, so that 
 the influence of the difference between the cross sections is reduced in 
 proportion to the intensity contrast of both source terms. Hence we assume, 
 for our purposes, that $\sigma_{\rm distant}\,\simeq\,\sigma_{\rm 
 halo}\,\simeq\,\sigma_{\rm extragal}$ towards high $|b|$.
 
 We assume thus that $I_{\rm LHB}$\,=\,const. within each field examined. 
 Deviations from this assumption will be revealed by failures of our model 
 to account for the observed soft X-ray emission. $I_{\rm distant}$ is 
 dominated by the $I_{\rm halo}$ term because $I_{\rm halo} > I_{\rm 
 extragal}$ towards high-$|b|$ directions. To separate $I_{\rm halo}$ and 
 $I_{\rm extragal}$, we would need supplementary {\it ROSAT\/} PSPC pointed 
 data (see Barber et al. 1996; Cui et al. 1996; and our discussion in 
 Section 5.1). We thus arrive at the simplified radiation-transfer equation
 \begin{equation}
 I_{\rm SXRB}\,=\,I_{\rm LHB}\,+\,I_{\rm distant}\cdot e^{-\sigma_{\rm distant}
 \cdot\ N_{\rm HI}({\rm total)}}
 \end{equation}
 This equation is of the form earlier studied by Marshall \& Clark (1984).
 In the following we show that Eq.\,(2) represents the observed situation well.

 \subsection{Evaluation of the radiation-transfer equation}
 
 \subsubsection{The general approach}
 
 We evaluated the SXRB radiation-transfer equation (Eq.\,2) using several 
 different methods. Table 1 lists the $N_{\rm HI}$ range for each field. 
 Traced by $N_{\rm HI}$, we evaluate $\sigma_{\rm distant}$ and the 
 corresponding attenuation of $I_{\rm distant}$. A standard method (see e.g. 
 Herbstmeier et al. 1995) involves
 fitting Eq. (2) to the data, plotted in the form of a scatter diagram
 of observed SXRB count rate versus total $N_{\rm HI}$. The disadvantage of 
 such a method is that it neglects the positional information of the
 data.
 
 An alternative method was introduced by Kerp et al. (1996), who questioned 
 the assumption that all of the \hi is located between the observer and the 
 distant X-ray sources; the SXRB/$N_{\rm HI}$ relation might depend on the 
 kinematic range of integration entering $N_{\rm HI}$. They evaluated the 
 modelled SXRB intensity distribution according to Eq. (2) for each image 
 pixel. Hence they determined the deviation between the observed and the 
 modelled SXRB intensity distribution, giving a measure of the degree of 
 correlation or anti-correlation of observed and modelled SXRB images. By 
 averaging the individual deviation values of the image pixels across the 
 entire field, they calculated the brightness of the source terms in Eq. 
 (2). The intensities $I_{\rm LHB}$ and $I_{\rm distant}$ were tuned to 
 minimize the difference between both images. This method accounts for the 
 location of the X-ray absorbing clouds within
 the field and directly reveals the areas where the
 X-ray data significantly deviate from the modelled mean intensity
 values.
 
 Here, we optimize the method of Kerp et al.
 (1996) with respect to evaluation of the derived count rate
 of the $I_{\rm distant}$ component.
 We calculated the optimal $I_{\rm distant}$
 value using Eq. (2) individually for each image pixel. For instance, if the 
 distant X-ray source is patchy or if the distribution of $N_{\rm HI}$ does 
 not correctly trace the amount of X-ray absorbing matter (perhaps due to 
 neglecting the existence of ${\rm H_2}$ and ${\rm H^+}$), then a very 
 patchy modelled SXRB intensity pattern would have followed, whereas, in fact, 
 it was determined as quite constant.
 
 \subsubsection{First results}
 Fig.\,\ref{fig:hvcc-low} illustrates our results, comparing, for one of our 
 fields, the SXRB distribution observed by the {\it ROSAT\/} PSPC with the 
 modelled situation. In order to calculate this modelled map, we determined a 
 constant $I_{\rm distant}$ intensity level across the entire field. In our 
 procedure we let a constant $I_{\rm distant}$ X-ray background intensity 
 penetrate through the absorbing neutral interstellar medium -- Fig.\,2c 
 shows the $N_{\rm HI}$ distribution as tracing absorption at $|v_{\rm 
 LSR}|\,\leq\,100\,{\rm km\,s^{-1}}$ -- and add the $I_{\rm LHB}$ emission, 
 also assumed to be constant, to this attenuated SXRB map. We tuned both 
constant X-ray 
 source intensity levels of Eq.\,(2) in order to obtain the best fit to the 
 observations. 

To quantify this result, we tested the hypothesis that the differences 
between the observed and modelled intensity distributions are statistical 
deviations and not uncertainties introduced by the modelling of the X-ray 
data.  The observed minus modelled X-ray intensity distribution was binned 
into a histogram (100 bins) showing the frequency of the deviation versus 
the deviation value.  The histogram was quantitatively compared with a 
Gaussian distribution using a $\chi^2$ test.  We found $\chi^2 = 67$, well 
below the acceptable value of $\chi^2=120$ for 96 degrees of freedom, and
a rejection threshold of 0.05.  
The hypothesis that both distributions are significantly different has to 
be rejected. This confirms that our approach of assuming constant $I_{\rm 
LHB}$ and $I_{\rm distant}$ matches the observed situation well.  
Additionally, this finding confirms that \hi is the best tracer of the 
photoelectric absorption and that H$^+$ as well as H$_2$ influence the soft 
X-ray radiation transfer on a much lower level compared to \hi. Thus, we 
conclude that $I_{\rm distant}$ can be 
 approximated well by an intensity which is constant across the entire 
 field: the distant soft X-ray background radiation is not patchy on angular 
 scales of some tens of degrees. This finding was verified for all 
 analyzed fields, distributed across the sky. The absolute value of $I_{\rm 
 distant}$ varies significantly, however, between the individual fields. 
 Because $I_{\rm extragal}$ is plausibly constant across the entire sky, the 
 large-scale variation of $I_{\rm distant}$ is entirely attributed to 
 $I_{\rm halo}$. This will be discussed in detail in Sect.\,5.1.
 
 \subsubsection{Interpretation of the results}
 
 Following the procedures described below, we scaled the intensity
 of a constant-intensity X-ray background source beyond the entire $N_{\rm 
 HI}$ contribution shown in Fig.\,2c. This yielded the image of the modelled 
 SXRB intensity distribution shown in Fig.\,\ref{fig:hvcc-low}b. In 
 Fig.\,\ref{fig:hvcc-low}a, we superposed, as contour lines, the deviations 
 between the observed and the modelled SXRB intensity distribution, starting 
 with the 4-$\sigma$ level and increasing in steps of 2$\sigma$.
 Dashed lines indicate areas where the modelled SXRB intensity is
 too bright, or where we missed additional X-ray absorbers not
 traced by the \hi radiation; solid lines mark areas where {\it ROSAT\/} 
 detected more radiation than expected by the \hi data. At these positions, 
 we have either overestimated the amount of absorbing matter or we are 
 observing true excess X-ray emission. This excess corresponds to some 25\% 
 of the total SXRB intensity.
 In general, an underestimate of the amount of matter attenuating the X-rays 
 is more likely than an overestimate, because neither H$_2$ nor H$^+$ is 
 represented by the 21-cm tracer.  Thus, it seems likely that the dashed 
 contours indicate the presence of additional absorbing matter, 
 but that the solid contours indicate X-rays in excess of the average.
 
 \subsubsection{Evaluation of $I_{\rm distant}$}
 
 We evaluated the level of the modelled constant distant X-ray source 
 intensity using three additional methods.
 First, we averaged the X-ray halo intensities across the entire map over 
 areas of equal $N_{\rm HI}$ in bins of $\Delta N_{\rm 
 HI}\,=\,1\cdot10^{19}\,{\rm cm^{-2}}$. This
 yielded the dependence of the X-ray halo intensities
 on the amount of absorbing \hi shown in Fig.\,\ref{fig:idist_nh}.
 The slope of the dependence is a function of the assumed $I_{\rm LHB}$
 count rate. If the $I_{\rm LHB}$ count rate is underestimated, we
 obtain a correlation of X-ray halo intensity with $N_{\rm
 HI}$; in case of an overestimate of $I_{\rm LHB}$, we obtain an 
 anti-correlation.
 We tuned the $I_{\rm LHB}$ value such that the dependence is minimized. 
 This alignment corresponds to the assumption that the $\frac{1}{4}$\,keV 
 radiation is independent of the amount of \hi along the line of sight.
 Such is certainly not the case for specific areas of the galactic sky. For 
 example, towards the North Polar Spur (Egger \& Aschenbach 1995) the X-ray 
 intensity is {\em not} distributed independently from the $N_{\rm HI}$ 
 structure. 
 
 Second, we averaged both the $\frac{1}{4}$\,keV and the $N_{\rm HI}$ data
 over $l$ and $b$, respectively, and compared these mean observed intensity 
 values with the model. This method allows searching for systematic 
 uncertainties introduced by the modelling of the X-ray data. We tested the 
 hypothesis that areas of the sky with 
 the same $N_{\rm HI}$ values correspond to unique 
 $I_{\rm distant}$ and $I_{\rm LHB}$ values, within the uncertainties of the 
 X-ray data.
 We evaluated the dependence of the source terms in Eq. (2) on the
 galactic $l$ and $b$ profiles. The derived 
 values for $I_{\rm LHB}$ and $I_{\rm distant}$ agree with those
 calculated by the first method.
 Fig.\,\ref{fig:hvcc-low}f shows the dependence on $l$ and $b$
 of the soft X-ray radiation-transfer equation solved with
 the same intensity values for the LHB and the galactic halo
 plasma as used to derive panel (b) of Fig.\,\ref{fig:hvcc-low}. There are 
 no significant large-scale differences between the $l$ and $b$ 
 distributions of the observed SXRB radiation and the modelled X-ray 
 intensity derived from the $N_{\rm HI}$ distribution. This indicates that 
 the distant soft X-ray emission is constant, within the statistical
 limitations of the X-ray data, across each field.
 
 Third, we averaged observed SXRB count rates with a given 
 $N_{\rm HI}$ in steps of $\Delta
 N_{\rm HI}\,=\,1\cdot10^{19}{\rm cm^{-2}}$ and plotted a simple
 scatter diagram of $I_{\rm SXRB}$ versus $N_{\rm HI}$. 
 This method is sensitive to the choice of the source term parameters 
 in Eq. (2), and would reveal erroneous model 
 parameters.
 
 Thus, we confirmed the validity of the soft X-ray
 radiation-transfer solution using three independent methods.
 The second and, even more so, the third, method suffers from neglect of the 
 positional information in the {\it ROSAT\/} maps. But they show 
 that the $I_{\rm distant}$ values returned are consistent with those of the 
 first method, which does account for the positional information. This 
 indicates that the $I_{\rm distant}$ source term is, within the statistical 
 limitations of the X-ray data, constant on angular scales of several tens 
 of degrees.
 
 We considered the uncertainties of the individual soft X-ray source terms 
 by varying $I_{\rm LHB}$ or $I_{\rm distant}$ independently in a way that 
 the modelled and observed intensities fit within the statistical 
 uncertainties of the data. Because the quantities are field-averaged, the 
 corresponding uncertainties are low. For the local X-ray emission, $\Delta 
 I_{\rm LHB}\,\simeq\,0.5\cdot 10^{-4}\,{\rm cts\,s^{-1}\,arcmin^{-2}}$; 
 $\Delta I_{\rm distant}\,\simeq\,3.0\,-\,7.0\cdot 10^{-4}\,{\rm 
 cts\,s^{-1}\,arcmin^{-2}}$, depending on the averaged $N_{\rm HI}$ value 
 across the field, and thus typically an order of magnitude higher than the 
 LHB plasma.
 \subsection{X-ray absorption traced by \hi}
 
 \subsubsection{Velocity information in the \hi data}
 
 Above, we described our investigation of the $I_{\rm LHB}$ and $I_{\rm 
 distant}$ source terms of Eq. (2). Now, we show how we determined the 
 amount of \hi absorbing the soft X-rays.
 The velocity information contained in the \hi data gives an additional free
 parameter in Eq. (2). We can integrate the \hi brightness temperatures over 
 different velocity intervals, introducing a kinematic unravelling which may 
 indicate also a spatial separation. Three separate velocity regimes
 are commonly, albeit somewhat arbitrarily, distinguished in the literature, 
 namely as low-velocity (LV: $|v_{\rm LSR}|\,\leq\,25\,{\rm
 km\,s^{-1}}$), intermediate-velocity (IV: $25\,{\rm
 km\,s^{-1}}\,\leq\,|v_{\rm LSR}|\,\leq\,90\,{\rm km\,s^{-1}}$),
 and high-velocity (HV: $|v_{\rm LSR}|\,\geq\,90\,{\rm
 km\,s^{-1}}$). The low-velocity regime not only samples all of the 
 higher-$|b|$ \hi which belongs to the conventional galactic disk, it 
 includes most of the \hi which corresponds to the warm diffuse \hi 
 layer (e.g. Dickey \& Lockman 1990, also denoted as {\it warm neutral medium\/}, WNM) as well.
 If we integrate the \hi spectra over the low-velocity regime, we neglect 
 some 10\% of the total amount of \hi distributed across the 
 field, although this percentage varies from region to region. In some 
 regions, there is as much emission from \hi gas at extreme velocities 
 as from LV matter; towards these lines of sight it is not feasible to 
 evaluate the soft X-ray radiation transfer only with the low-velocity 
 $N_{\rm HI}$. The Draco cloud (Herbstmeier et al. 1996) is an example of an 
 IVC dominating $N_{\rm HI}$; in addition, it contains significant amounts 
 of molecular matter. Finally, HVCs may also absorb the distant SXRB source 
 radiation (see Herbstmeier et al. 1995).

 \subsubsection{Dependence of $I_{\rm distant}$ and $I_{\rm LHB}$ on the 
 $N_{\rm HI}$ velocity interval}
 
 To test whether the choice of $N_{\rm HI}$ velocity interval 
 reveals a kinematic unravelling of the source of the SXRB, we integrated the 
 \hi emission separately over the LV range 
 $|v_{\rm LSR}|\,\leq\,25\,{\rm km\,s^{-1}}$,
 and over two wider velocity ranges
 $|v_{\rm LSR}|\,\leq\,50\,{\rm 
 km\,s^{-1}}$ and 
 $|v_{\rm LSR}|\,\leq\,100\,{\rm km\,s^{-1}}$
Towards high galactic latitudes the latter range encompases all interstellar
 gas except the
HVCs.
 The histograms in Fig.\,\ref{fig:idist_nh} represent $I_{\rm distant}\,=
 \,(I_{\rm SXRB}\,-\,I_{\rm LHB})\,e^{\sigma_{\rm distant}\cdot
 N_{\rm HI}{\rm (total)}}$ as a function of $N_{\rm \hi}$.
 Within the uncertainties of the histogram data points (the error bar in Fig.\,
 \ref{fig:idist_nh}), $I_{\rm distant}$ can be considered a constant 
 across the $N_{\rm \hi}$ range of $0.5\cdot10^{19}\,{\rm cm^{-2}}$\,to\,
 $8.0\cdot10^{20}\,{\rm cm^{-2}}$.
 The horizontal solid line represents our field-averaged best-fit
 value of
 $I_{\rm distant}$, while the
 dashed lines mark the uncertainties of this best-fit value.
 Taking into account the uncertainties of both, the data and the modelling,
 the assumption of a constant $I_{\rm distant}$ is justified.
 With Fig.\,\ref{fig:idist_nh} we can also constrain the expected intensity
 variation of the $I_{\rm LHB}$ source term,
 because to evaluate $I_{\rm distant}$ as a function of 
 $N_{\rm \hi}$ we {\it a priori\/} assumed $I_{\rm LHB}$\,=\,const.
 Consequently, our finding $I_{\rm distant}$\,=\,const. implies $I_{\rm LHB}$\,=
 \,const. within the uncertainties of the analysis.
 
 The three histograms in Fig.\,\ref{fig:idist_nh} show that the functional
 dependence of $I_{\rm distant}$ on $N_{\rm \hi}$ is independent of
 the extent of the velocity range used to evalute $N_{\rm \hi}$.
 
 However, the mean level of 
 $I_{\rm distant}$ increases proportionally to the extent of the integration 
 range of $N_{\rm \hi}$. Nevertheless, all data points of the
 three histograms are within the uncertainty range of the modelled 
 $I_{\rm distant}$
 intensity level. We conclude that the 
 WNM in the Galaxy determines the mean intensity level of the
 distant soft X-ray background radiation.
 Towards high galactic latitudes, the WNM best represents the physical state
 of the major fraction of the interstellar matter. Accordingly, the \hi
 belonging to the WNM traces the amount of soft X-ray absorbing matter,
 and determines the mean intensity level of the distant diffuse X-ray radiation.
 The bulk of the  WNM is already enclosed in the
 velocity bracket $|v_{\rm LSR}|\,\leq\,25\,{\rm km\,s^{-1}}$ 
 (Dickey \& Lockman 1990). Accordingly, the additional $N_{\rm \hi}$ at more
 extreme velocities increases the mean $I_{\rm distant}$ level, but does not
 change significantly the functional dependence of $I_{\rm distant}$ on
 $N_{\rm \hi}$.

 The discussion above implies that our modelling of the {\it ROSAT\/} X-ray
 data can be well approximated by constant $I_{\rm LHB}$ and $I_{\rm distant}$
 source terms across the extent of the fields of interest (question 1 of Sect.\,
 3). The mean $I_{\rm distant}$ X-ray intensity level is determined by the
 distribution of the WNM gas. The more extreme
 velocity ranges represent, on the average, only
 a minor fraction of the total interstellar gas. Accordingly, the $|v_{\rm LSR}|
 \,\leq\,25\,{\rm km\,s^{-1}}$ is sufficient to determine the $N_{\rm \hi}$
 responsible for the attenuation of the distant diffuse X-ray sources (question
 3 of Sect\,3).
 However, our aim is to search for soft X-ray enhancements of HVCs, with respect
 to the finding shown in Fig.\,\ref{fig:idist_nh} the $|v_{\rm LSR}|\,\leq\,100
 \,{\rm km\,s^{-1}}$ yields the highest $I_{\rm distant}$ intensity level.
 This attributes a maximum of the diffuse X-ray emission to $I_{\rm distant}$
 and introduces a systematic bias in our analysis for the {\em non--detection}
 of excess X-ray emission associated with HVCs.
 
 \section{Individual HVC complexes}
 
 To investigate whether soft X-ray enhancements
 are associated with HVCs, we {\em excluded} the HVC velocity regime from the
 velocity range used to determine the absorbing $N_{\rm HI}$, in particular
 we integrated $N_{\rm HI}$ over $|v_{\rm LSR}|\,\leq\,100\,{\rm km\,s^{-1}}$.
 This exclusion
 introduces the brightest modelled SXRB intensity just at the
 positions of the HVCs and thus biases our analysis against detection of 
 soft X-ray enhancements with HVCs, because we evaluate observed minus modelled
 X-ray intensity distribution only. We now evaluate the solutions of Eq. (2) 
 with $N_{\rm HI}$ determined over the more extreme velocity interval 
 $|v_{\rm LSR}|\,\leq\,100\,{\rm km\,s^{-1}}$, searching for 
 soft X-ray correlations or anti-correlations with HVCs.

 \subsection{The HVC complex C}
 
 \subsubsection{Complex C at lower $l$ and $b$, and parts of complex D}
 
 Kerp et al. (1996) investigated the X-ray intensity distribution towards 
 complex C at $34\degr\,\leq\,l\,\leq\,86\degr$, 
 $33\degr\,\leq\,b\,\leq\,79\degr$. Here, we discuss parts of complex C, 
 weaker in \hi emission, at lower $b$. Figure\,\ref{fig:hvcc-low}a shows the 
 {\it ROSAT\/} PSPC data from the lower-$b$ part of complex C. Panel (b) 
 shows the modelled SXRB intensity distribution, assuming a constant SXRB 
 source intensity across the field. We derived the intensity of the LHB, 
 $I_{\rm LHB}\,=\,(2.8\pm 0.5)\cdot10^{-4}\,{\rm cts\,s^{-1}\,arcmin^{-2}}$,
 and of the distant X-ray source, $I_{\rm distant}\,=\,(25 \pm 4
 )\cdot10^{-4}\,{\rm cts\,s^{-1}\,arcmin^{-2}}$.
 Both X-ray images in Fig.\,\ref{fig:hvcc-low}a and b are scaled similarly. 
 A statistical evaluation of the similarity between the observed and modelled
 X-ray map gives $\chi^2\,=\,67\,<\,{\chi^2}_{0.05}\,=\,120$. This confirms
 that also statistically the observed and modelled X-ray intensity
 distributions match.
 In Fig.\,\ref{fig:hvcc-low}a we superposed, as contours, the deviations 
 between the observed and the modelled SXRB 
 distributions, starting with the 4-$\sigma$ contour level. Dashed contours 
 indicate where the modelled SXRB intensity is brighter than observed 
 ; solid contours enclose regions where the modelled SXRB intensity is 
 weaker than observed.
 
 The dashed contours do not enclose the positions of individual HVCs (see 
 Fig.\,\ref{fig:hvcc-low}d), indicating that we do not detect soft X-ray 
 shadows of HVCs at this significance level. It is more likely that 
 these dashed contour lines of X-ray shadows indicate cloud structure within
 the LHB.
 As mentioned in Sect.\,3.3.1, 
 the effective photoelectric absorption cross section of the LHB plasma is 
 the largest of all three cross sections: an \hi cloud within the LHB will 
 cause a deeper soft X-ray shadow than when the same cloud were located 
 outside the LHB. In consequence, if predicted soft X-ray emission is weaker 
 than observed, one first has to check for the existence of a cloud within 
 the LHB. The dashed contours in Fig.\,\ref{fig:hvcc-low}a show a patchy 
 distribution; a large area of weaker X-ray emission is located at 
 $l\,=\,70\degr\,-\,85\degr$, $b\,\geq\,30\degr$. Located close to the 
 dashed contours is an elongated \hi filament, part of a much more extended
 local \hi structure (Wennmacher et al. 1998, Kerp \& Pietz 1998).
 An $N_{\rm \hi}$ maximum of this structure associated with a filament,
 denoted as LVC\,88+36--2, was studied by Wennmacher et al. (1992).
 Kerp et al. (1993) detected a strong soft X-ray 
 absorption feature associated with LVC\,88+36--2 in pointed {\it ROSAT\/} 
 PSPC data and confirmed that the filament is embedded within the LHB. Thus, 
 the dashed contours indicate, most likely, local $N_{\rm HI}$ maxima of an
 extended \hi structure within the LHB (see Kerp \& Pietz 1998).
 
 A second region of low observed SXRB emission, at 
 $l\,=\,45\degr\,-\,65\degr$ and $b\,\geq\,40\degr$, is not associated with 
 a previously identified local \hi structure. As mentioned in Sect.\,3.4.3, 
 an underestimate on the amount of X-ray absorbing matter is more likely 
 than an overestimate, because \hi emission traces neither molecular 
nor 
 ionized gas. The dashed contours may indicate an additional 
 absorber, 
 either located outside of the LHB (and thus only 
 attenuating the $I_{\rm distant}$ term) or within the local bubble.
 In the former case, we miss $\Delta N_{\rm HI}\,\simeq\,4\cdot 
 10^{20}\,{\rm cm^{-2}}$ as an absorber; in the latter case, 
 $\Delta N_{\rm HI}\,\simeq\,1\cdot 10^{20}\,{\rm cm^{-2}}$.  This 
 difference in absorbing $N_{\rm HI}$ between both model assumptions follows 
 from different amplitudes of the near and distant photoelectric absorption 
 cross sections (see Fig.\,\ref{fig:cross}). Consequently, the SXRB minimum 
 is more likely due to a local cloud than to a cloud of higher $N_{\rm 
 HI}$ beyond the local bubble. Hartmann et al. (1998) detected no molecular 
 gas in this direction, although such gas might be anticipated for a cloud 
 outside of the LHB with such a high \hi density. A further investigation 
 of the Leiden/Dwingeloo data reveals an \hi minimum at $v_{\rm LSR} \approx 
 -2 {\rm km\,s^{-1}}$, suggesting that some of the local \hi may have been 
 ionized and not quantitatively traced by the distribution of $N_{\rm \hi}$.
 The distance to the absorber thus remains uncertain.
 
 The solid contours in Fig.\,\ref{fig:hvcc-low}d enclose an HVC catalogued 
 as \#182 by Wakker \& van Woerden (1991), at $v_{\rm LSR}\,=\,-190\,{\rm 
 km\,s^{-1}}$, and attributed to HVC complex D. Our analysis suggests an 
 excess soft X-ray emission with a significance level greater than 
 4$\sigma$. The solid contours also enclose nearby regions of 
 intermediate-velocity gas ($-75\,{\rm km\,s^{-1}} \leq v_{\rm LSR} \leq 
 -25\,{\rm km\,s^{-1}}$, Fig.\,\ref{fig:hvcc-low}e), implying that IVCs may 
 also be associated with the enhanced X-ray emission.
 In this particular case, where both HVC and IVC gas appear along the same 
 lines of sight, we cannot determine whether the HVCs or the IVCs are the 
 sources of the excess soft X-ray emission.
 
 Finally, we analyzed the variation of the modelled 
 (Fig.\,\ref{fig:hvcc-low}b) and observed (Fig.\,\ref{fig:hvcc-low}a) SXRB 
 emission, as averaged over $l$ and $b$. We solved the radiation-transfer 
 equation independently for these averaged distributions. 
 Figure\,\ref{fig:hvcc-low}f shows the observed and modelled SXRB intensity 
 profiles averaged in $l$ and $b$. The modelled SXRB intensity profile 
 (solid line) fits the {\it ROSAT\/} observation (dots) well. This shows 
 that the dominant part of the soft X-ray attenuation is traced by \hi, and 
 that small-scale ($0\fdg8$) as well as large-scale ($\sim\,30\degr$) 
 intensity variations of the SXRB can be explained by photoelectric 
 absorption. This result justifies again our assumption that $I_{\rm LHB}$\,=
 \,const. and $I_{\rm distant}$\,=\,const. (see Sect.\,3.3) across each field. 

 \subsubsection{Complex C at higher $l$ and $b$}
 
 Figure\,\ref{fig:hvcc-high}a shows the {\it ROSAT\/} $\frac{1}{4}$\,keV 
 map of complex C between $99\degr\,\leq\,l\,\leq\,166\degr$, 
 $12\degr\,\leq\,b\,\leq\,74\degr$. The field also includes much of HVC 
 complex A as well as the high-velocity filament which connects HVC complex 
 C with A (Wakker \& van Woerden 1991). 
 The map covers such a large range in $l$ ($\simeq\,67\degr$) and $b$ 
 ($\simeq\,62\degr$) that the $N_{\rm HI}$ distribution varies appreciably 
 across the field. This yields the opportunity to study the variation of the 
 SXRB source intensity distribution with galactic latitude. In the upper left 
 of Fig.\,\ref{fig:hvcc-high}a, strong soft X-ray attenuation by the neutral 
 matter associated with the North Celestial Pole Loop (Meyerdierks et al. 
 1991) is visible (see also Fig.\,\ref{fig:hvcc-high}c, $l\,\sim\,135\degr$, 
 $b\,\sim\,35\degr$). Significant amounts of
 molecular material are found near this structure (Heithausen et al. 1993), 
 for instance in the Polaris Flare ($l\,\sim\,125\degr$, $b\,\sim\,30\degr$; 
 Heithausen \& Thaddeus 1990). Towards the Polaris Flare, the 
 Leiden/Dwingeloo data show a maximum of $N_{\rm HI}\,\simeq\,9\cdot 
 10^{20}{\rm cm^{-2}}$ (Fig.\,\ref{fig:hvcc-high}c). The Lockman et al. 
 (1986) area of minimum $N_{\rm HI}$ ($l\,\sim\,152\degr$, 
 $b\,\sim\,52\degr$) is located at the other end of the field. The data show 
 a ratio $N_{\rm HI\,max}/N_{\rm HI\,min}\,=\,25$ in the absorbing column 
 densities.
 
 We evaluated the X-ray source terms using all three methods described in 
 Sect.\,3.4.4 and found $I_{\rm LHB}\,=\,(3.5\,\pm\,0.5) \cdot 10^{-4}\,{\rm 
 cts\,s^{-1}arcmin^{-2}}$ and $I_{\rm distant}\,=\,(16\,\pm\,3) \cdot 
 10^{-4}\,{\rm cts\,s^{-1}arcmin^{-2}}$.
 The $\chi^2$-test of the observed and modelled X-ray map indicates that
 $\chi^2\,=\,170\,>\,{\chi^2}_{0.05}\,=\,120$.
 The differences between the observed and modelled map are significant. Most
 probably, the structure of the interstellar medium covered by the field of 
 interest is much too inhomogenious to be fitted by our simple approach.
 However, Fig.\,\ref{fig:hvcc-high}f shows that the modelling of the X-ray
 data fits the overall SXRB intensity distribution well, especially if we
 take the bright X-ray enhancements around $b\,\sim\,50\degr$ into
 consideration.
 However,to distinguish between excess emission areas and
 large scale intensity variations of $I_{\rm LHB}$ and $I_{\rm distant}$,
 we restrict our interpretation of the X-ray deviations to high
 galactic latitudes ($b\,\geq\,35\degr$) and to peak deviations more
 significant than $5\sigma$.
 In Fig.\,\ref{fig:hvcc-high}a, most 
 of the contours are oriented parallel to $b\,\sim\,50\degr$. Nearby are the 
 main parts of HVC complex C (see Fig.\,\ref{fig:hvcc-high}d) and the 
 Lockman et al. window, which is enclosed by dashed 
 contours.  In this region, our $I_{\rm LHB}$ value is lower by a factor of 
 two than the value given by Snowden et al. (1994b, 1998), while $I_{\rm 
distant}$ is 
 higher by about the same factor. To investigate this discrepancy 
 (see Freyberg 1997), we extracted the Lockman Window data from our map and 
 evaluated the radiation transfer equation in this area once again, 
 restricting our analysis to a region of $12\degr$ in extent in both $l$ and 
 $b$. We derived $I'_{\rm LHB}\,=\,(6.0\,\pm\,0.5) \cdot 10^{-4}\,{\rm 
 cts\,s^{-1}arcmin^{-2}}$ and $I'_{\rm distant}\,=\,(7\,\pm\,3) \cdot 
 10^{-4}\,{\rm cts\,s^{-1}arcmin^{-2}}$, applying only the first method 
 described in Sect.\,3.4.4.
 Using the second and third methods, we find values which, although in 
 closer agreement with Snowden et al. (1994b), 
 do not fit the averaged $l$ and $b$ intensity profiles. 
 Moreover, if we extrapolate the $I'_{\rm LHB/distant}$ values to the 
 data shown in Fig.\,\ref{fig:hvcc-high}a, we fail 
 to reproduce the observations. In contrast, the first-method values 
 ($I_{\rm LHB/distant}$) do fit the Lockman Window region in the averaged 
 $l$ and $b$ profiles (see Fig.\,\ref{fig:hvcc-high}f). 
 This shows that a solution of the 
 radiation-transfer equation demands determination of 
 $I_{\rm distant}$ over areas large enough not to be biased by local 
 events.
 
 The solid contours roughly trace some of the brighter parts of HVC complex 
 C, suggesting that these bright HVCs, in addition to other parts of complex 
 C (Kerp et al. 1996), are associated with excess soft X-ray emission.  The 
 Pietz et al. (1996) \hi ``velocity-bridges'' suggest the interaction of 
 some HVC matter with the conventional-velocity regime.
 The velocity bridges VB\,112\,+48 and 
 VB\,115\,+\,47, both at the high-$b$ end of complex C, 
 are enclosed by 5-$\sigma$ contours. VB\,112\,+57.5 is an area of soft 
 X-ray radiation enhanced to the 11-$\sigma$ level. VB\,111\,+\,35 and 
 VB\,133\,+\,55 were not detected as enhancements in the $\frac{1}{4}$\,keV 
 {\it ROSAT\/} data. If we add the four velocity bridges already found to be 
 X-ray bright by Kerp et al. (1996) using similar methods, we find that 7 of 
 11 bridges are located close to soft X-ray enhancements. The velocity 
 bridges span the range of conventional velocities to those of the HVCs,
 and thus their association with enhanced X-ray emission does not, in 
 itself, distinguish between an HVC or an IVC connection (see 
 Fig.\,\ref{fig:hvcc-high}e, and Sect.\,5.2). 
 
 The suggestion that velocity bridges are associated with a distortion of 
 the velocity field due to an HVC, requires disproving that the bridges are 
 different from normal IVC structures. The large distance to the HVCs 
 (several kpc for the nearest, Wakker \& van Woerden 1997, and possibly 
 hundreds of kpc for many HVCs, Blitz et al. 1998) make it unlikely that 
 HVCs are physically linked to those IVCs which are carriers of dust.
 Much of the intermediate-velocity \hi is associated with dust 
 cirri (Deul \& Burton 1990), and is therefore likely to be rather local, 
 rarely extending to $z$-heights of more than 150 pc.  The velocity regime 
 of cirrus-carrying IVCs, however, is frequently trespassed upon by HVCs: 
 the crossing of the Magellanic Stream from positive to negative velocities 
 is a case in point. HVC gas trespassing on lower velocities will have a 
 different chemical composition from the dust-carrying IVCs. In Sect.\,5.2 
 we will discuss this point for HVC complex C in more detail. 
 
 In the special case of a wide extent in galactic latitude, it is 
 interesting to study the observed and modelled SXRB
 intensity variations against $l$ and $b$, as shown in 
 Fig.\,\ref{fig:hvcc-high}f. Again, the solid line marks the modelled 
 intensity profile based only on \hi data. The $b$-variations show 
 quantitative agreement between observationed and modelled values, deviating 
 only close to $b\,\simeq\,40\degr$, at the location of the 
 {North Celestial Pole Loop} (region of highest opacity), and above 
 $b\,\geq\,70\degr$. These deviations are significant: the error-bars 
 correspond to the 3-$\sigma$ level. Most likely, we observe an intensity 
 variation of $I_{\rm LHB}$ proportional to increasing $b$. 
 Fig.\,\ref{fig:hvcc-high}f shows that the X-ray intensity variation is 
 correctly predicted by the modelled SXRB intensity distribution, but 
 starting at $b\,\geq\,60\degr$, the modelled SXRB intensity deviates 
 increasingly from the observed one. Towards these high-$b$ regions 
 we may predominantly observe the local interstellar medium, and consequently a 
 larger extent of the local X-ray emitting region. Finally, we note that the 
 modelled SXRB longitude profile closely matches the observed one 
 for $l\,\ga\,130\degr$. This position coincides with the border of the 
 X-ray enhancements associated with HVC complex C.

 \subsection{The HVC complex GCN}

 The mean $N_{\rm HI}$ towards the galactic center HVC complex GCN 
 ($18\degr \leq l \leq 73\degr$, $-52\degr \leq b \leq -15\degr$) is 
 significantly higher than towards the other regions discussed here. The 
 field displays the complex \hi column density structure within the range $|v_{\rm 
 LSR}| \leq 100\,{\rm km\,s^{-1}}$ (Fig.\,\ref{fig:gcn}c). Solving of the 
 radiation-transfer 
 equation gives $I_{\rm LHB}\,=\,(2.3\,\pm\,0.5)\cdot
 10^{-4}\,{\rm cts\,s^{-1}arcmin^{-2}}$ and $I_{\rm
 distant}\,=\,(30\,\pm\,7)\cdot 10^{-4}{\rm
 cts\,s^{-1}arcmin^{-2}}$.
 The $\chi^2$ test of the observed and modelled data gives $\chi^2\,=\,59\,<\,
 {\chi^2}_{0.05}\,=\,120$.
 Fig.\,\ref{fig:gcn} shows the {\it ROSAT\/} data (panel a) 
 our solution of Eq. (2) (panel b) using the Leiden/Dwingeloo \hi data.
 This field shows well-defined large-scale X-ray intensity gradients in the 
 {\it ROSAT\/} data which are reproduced by our solution of Eq.\,(2), 
 confirming that the intensity variations are dominated by photoelectric 
 absorption effects.
 
 The distant X-ray intensity ($I_{\rm distant}$) is quite high, and (within the 
 uncertainties) equal to the intensity value of the 
 lower end of HVC complex C (Sect. 4.1.1). Furthermore, the $I_{\rm LHB}$ intensity of 
 both areas agree. We note that both {\it ROSAT\/} areas cover a comparable 
 range in galactic coordinates, but refer to opposite galactic hemispheres, 
 which suggests that closer to the inner Galaxy, the northern and southern 
 galactic sky have approximately the same SXRB distant source intensity, and 
 that $I_{\rm distant}$ is not patchy across the individual fields. Thus, we 
 can consider $I_{\rm distant}$ as constant towards the same galactic 
 longitude range in both galactic hemispheres.
 
 The bright X-ray area, localed near in the center of 
 Fig.\,\ref{fig:gcn}a, shows excess soft X-ray emission enclosed by solid 
 contours.  The galaxy Mrk\,509 is marked by 
 the dot.  Sembach et al. (1995) used HST absorption-line measurements to 
 detect highly ionized high-velocity gas belonging to HVC complex GCN. They 
 attribute the source of the ionization to photoionization. Our {\it 
 ROSAT\/} data suggest, additionally, the presence of collisionally ionized 
 gas along the line of sight towards Mrk\,509. Sembach et al. 
 may have detected the cooler portion of the collisionally ionized plasma.  
{Figure\,\ref{fig:gcn}d shows the distribution of the GCN clouds across 
 the field.
 They are patchily distributed and have only low column densities of $N_{\rm 
 HI}\,\simeq\,5\cdot 10^{18}\,{\rm cm^{-2}}$. Very close by, some filaments 
 are found which belong to the HVC complex GCP; we can not distinguish whether the 
 excess emission originates in GCN or in GCP. Following 
 Sembach et al. (1995), we attribute the excess 
 emission to complex GCN.  Thus, in contrast to HVC complex C, where we 
 found a close positional correlation between neutral HVC gas and the X-ray 
 bright areas, the GCN complex allows no straightforward interpretation.  
 Blitz et al. (1998) include complex GCN amongst those suggested to be at 
 large, extragalactic distances.  If this is true, one must consider the 
 physical circumstances which would allow the presence of collisionally- 
 and photoionized gas associated with this complex.
 
 Figure\,\ref{fig:gcn}f shows the $l$ and $b$ profiles of
 the GCN maps. Again, the modelled SXRB intensity distribution fits the 
 observation, confirming that the areas of excess soft X-ray radiation are 
 well determined by the methods applied.
 \subsection{The HVC complex WA}
 
 HVC complex WA (Wannier et al. 1972; see also Wakker \& van Woerden 
 1991), roughly confined to the region $218\degr \leq l \leq 270\degr$, 
 $24\degr \leq b \leq 52\degr$, displays the positive velocities ($v_{\rm 
 LSR} \sim +150\,{\rm km\,s^{-1}}$) characteristic of most HVCs in this 
 general region of the sky. The radial velocity is, of course, only one 
 component of the velocity vector; the positive radial velocity does not 
 rule out, by itself, that the HVC could be colliding with the galactic 
 disk. Regarding the X-ray radiation transfer, it is interesting that the 
 HVC complex WA is located opposite the direction of HVC complex C, but also 
 in the northern sky.
 
 Figure\,\ref{fig:WA} shows the {\it ROSAT\/} $\frac{1}{4}$\,keV map 
 (panel a) and the modelled SXRB intensity (panel b). We derive an intensity of 
 $I_{\rm LHB}\,=\,(4.3\,\pm\,0.5)\cdot 10^{-4}\,{\rm
 cts\,s^{-1}arcmin^{-2}}$ for the LHB, and $I_{\rm 
 distant}\,=\,(13\,\pm\,4)\cdot 10^{-4}\,{\rm cts\,s^{-1}arcmin^{-2}}$ for 
 the distant X-rays. These LHB and distant X-ray count rates mark the extreme intensity
 values found in our sample of HVC complexes.
 Because of the short {\it ROSAT\/} integration times towards complex WA, 
 the observed and modelled SXRB maps show some deviations.
 The $\chi^2$-test of the maps reveal that $\chi^2\,=\,34\,<\,{\chi^2}_{0.05}\,
 =\,120$.
 Statistically, the modelled SXRB map fits the observed one well.
 However, only a few 4-$\sigma$ 
 contours are present in Fig.\,\ref{fig:WA}a).
 
 The SXRB radiation is locally weaker near $l\,\sim\,230\degr$, 
 $b\,\sim\,48\degr$ (4$\sigma$). Either we miss additional matter 
 attenuating the soft X-rays but not traced by $N_{\rm HI}$, or, more 
 likely, the photoelectric absorption cross section is locally larger due 
 to a cloud within the LHB plasma (see Fig.\,\ref{fig:cross}). At $v_{\rm 
 LSR} \approx 0\ {\rm km\,s^{-1}}$, \hi maps in the Leiden/Dwingeloo atlas 
 show a local deficiency of neutral gas, correlated with the contours given 
 in Fig. \ref{fig:WA}.
 
 Toward the general direction of HVC complex WA we identified soft 
 X-ray enhancements with known HVCs (Fig.\,\ref{fig:WA}d). The 
 4-$\sigma$ contour centered $l\,=\,258\degr$, $b\,=\,45\degr$ is 
 positionally associated with the HVC catalogued as \#66  by Wakker \& van 
 Woerden (1991). The solid contour near $l = 264\fdg8$, $b = 
 26\fdg5$, lies in between the HVCs catalogued as \#176 and \#162. (These HVCs 
 are within our WA field but Wakker \& van Woerden (1991) did not assign 
 them to complex WA.)  Because of the limited quality of the {\it ROSAT\/} 
 data and the possibility of residual systematic uncertainties, we do not 
 claim a firm detection of excess X-ray emission from the WA HVCs. 
 Fig.\,\ref{fig:WA}d shows the $N_{\rm HI}$ distribution of the HVCs towards 
 the field.
 The $N_{\rm HI}$ range of the HVCs displayed is only $5\cdot 10^{18}\,{\rm 
 cm^{-2}}\,\leq\,N_{\rm HI}\,\leq\,3\cdot 10^{19}\,{\rm cm^{-2}}$.} 
 Figure\,\ref{fig:WA}f shows the averaged SXRB intensity profiles derived 
 from the observed and modelled $\frac{1}{4}$\,keV SXRB data; within the 
 uncertainties of the data, the modelled SXRB variation fits the 
 observational data well.
 
 \section{Discussion}
 
 \subsection{The radiation transfer of $\frac{1}{4}$\,keV photons}
 
 We discuss here the accuracy of our solution of the radiation transfer in 
 confirming the detections of enhanced soft X-ray radiation close to HVCs, 
 and some general properties of the distant soft X-ray sources.
 The compelling similarity between the observed and the modelled
 SXRB intensity distributions, based only on \hi data, supports several 
 conclusions. It argues for a smooth intensity distribution of the SXRB 
 sources, at greater distances than the galactic \hi. Moreover, the 
 smoothness of the SXRB source distribution is emphasized by the success of 
 a constant intensity background distribution in fitting the {\it ROSAT\/} 
 data well across several tens of degrees as suggested by Pietz et al. 
 (1998b). This situation does not rule out that there may be large-scale 
 intensity gradients across the entire galactic sky. Also, the averaged 
 variations plotted against galactic $l$ and $b$ do not suggest that, there 
 are no intensity gradients in the SXRB, but they do indicate that within 
 the fields considered, the distant X-ray sources do not show significant 
 intensity variations.

 \begin{table}
 \caption[]{Summary of the derived $\frac{1}{4}$\,keV X-ray intensities, in 
 units of $10^{-4}\,{\rm cts\,s^{-1}arcmin^{-2}}$. The HVC complexes 
 investigated by Herbstmeier et al. (1995) and by Kerp et al. (1996) are 
 indicated by asterisks. The fields are ordered according to the angular 
 distance of each map center from the galactic center (see 
 Fig.\,\ref{fig:sxrb_long}).  The righthand column gives the $\chi^2$ value 
of the difference between the modelled and observed X-ray map.  Using a 
significance level of 0.05 the acceptable $\chi^2$ is 120 with 96 degrees 
of freedom.}
 \begin{flushleft}
 \begin{tabular}{llllll}
 \hline
 complex    & $l_{\rm c}$  & $b_{\rm c}$ & $I_{\rm LHB}$   & $I_{\rm 
distant}$ & $\chi^2$ \\
 \hline
 GCN         & $40\degr$    & $-32\degr$   & $2.3\,\pm\,0.5$ & $30\,\pm\,7$ 
& 59 \\
 C low, D    & $63\degr$    & $+32\degr$   & $2.8\,\pm\,0.5$ & $25\,\pm\,4$ 
& 67 \\
 C$^*$       & $94\degr$    & $+51\degr$   & $4.4\,\pm\,1.0$ & $18\,\pm\,3$ 
& 71 \\
 WA          & $247\degr$   & $+38\degr$   & $4.3\,\pm\,0.5$ & $13\,\pm\,4$ 
& 34 \\
 C high, A   & $132\degr$   & $+43\degr$   & $3.5\,\pm\,0.5$ & $16\,\pm\,3$ 
& 170 \\
 M$^*$       & $170\degr$   & $+60\degr$   & $\sim\,6.5$     & $\leq\,10$ & 
\\
 \hline
 \end{tabular}
 \end{flushleft}
 \end{table}
 
 Table 2 summarizes the derived intensities of the $I_{\rm LHB}$ and the 
 distant X-ray component, $I_{\rm distant}$, in order of increasing angular 
 distance of the map center from the galactic center.
 The variation of the galactic halo intensity noted in Table 2
 and plotted in Fig.\,\ref{fig:sxrb_long} suggests that towards the inner 
 Galaxy the distant soft X-ray
 source reaches a local maximum. Because we
 avoid the area of the North Polar Spur (Egger \& Aschenbach
 1995), this variation is probably due to the
 distant SXRB source component. Moreover, the distant SXRB source 
 intensities tend to decrease in the direction away from the galactic
 center (see Fig.\,\ref{fig:sxrb_long} and the discussion below).
 This variation with $l$ implies that we indeed observe galactic soft X-ray 
 emission, confirming the findings of Pietz et al. (1998a; 1998b).
 
 A similar intensity variation of the galactic X-ray halo component with $b$ 
 cannot be claimed from our data because all the X-ray maps analyzed are 
 at roughly the same latitude, near $|b|\,\sim\,35\degr$.
 Our data suggest, however, that the derived galactic X-ray halo
 intensity shows the same brightness in the northern and southern
 sky (Pietz et al. 1998b; Wang 1998). We note further that the derived LHB 
 intensities are proportional to the extent of the local cavity and
 in agreement with the shape of the LHB derived from absorption-line
 measurements (e.g. Egger et al. 1996). 
 
 The variations of the observed SXRB intensity, averaged over $l$ and $b$,
 indicate that, on large angular scales, the observed SXRB intensity 
 variation is determined, in detail, by the distribution of the absorbing 
 interstellar medium.  The similarity between the observed and modelled SXRB 
 maps shows that small-scale intensity variations $(\leq\,1\degr)$ of the observed SXRB can 
 also be attributed to photoelectric absorption.
 The soft X-ray absorption is well traced by 
 \hi in the velocity range $|v_{\rm LSR}|\,\leq\,100\,{\rm km\,s^{-1}}$.
 This range covers the conventional galactic gas as
 well as the IVCs. Because the chosen velocity range includes low-velocity 
 as well as intermediate-velocity \hi, and because in some cases the \hi 
 column from the IVC gas exceeds that of the conventional-velocity gas, 
 the X-rays have to originate beyond the IVCs studied by 
 Kuntz \& Danly (1996).
 From the soft X-ray shadow cast by HVC complex M (Herbstmeier et al. 
 1995), we conclude that at least a minor fraction of the galactic distant 
 X-ray emission originates at distances larger that of HVC complex 
 M. We conclude that nearly all galactic \hi 
 absorbs the X-ray
 halo radiation, because the vertical extent of the galactic
 \hi is entirely located within this distance
 range (Lockman \& Gehman 1991).
 
 Our analysis suggests that \hi alone predominantly traces the X-ray 
 absorption, because otherwise the modelled X-ray intensities would not fit 
 the observational data as well as they do.
 H$_2$ certainly absorbs the SXRB radiation along some lines of sight, but 
 is not diffusely distributed over scales of several tens of degrees, and is 
 rare at the higher galactic latitudes considered here (Magnani et al. 1997).
 Furthermore, the SXRB source intensity absorption traced by ${\rm H_2}$ occurs within 
 regions of high $N_{\rm HI}$, for instance as shown by our data towards the 
 Polaris Flare (Sect.\,4.1.2; Meyerdierks \& Heithausen 1996). Otherwise we 
 would have detected deep soft X-ray absorption features not traced by the 
 $N_{\rm HI}$ distribution, because $\sigma_{\rm X}({\rm 
 H_2})\,\geq\,2\cdot\sigma_{\rm X}({\rm HI})$.
 
 Soft X-ray absorption associated with diffusely distributed
 ionized hydrogen (Reynolds 1991) is also not obvious in
 our data. If the H$^+$ layer has a column density distribution
 similar to that of the \hi layer, we would
 anticipate a constant scaling factor for the brightness of
 the galactic halo X-ray component. On the other hand, if the distribution
 of ${\rm H^+}$ is patchy within the analyzed fields, its soft X-ray 
 absorbing column density would be about $\Delta N_{\rm H^+}\,\leq\,7\cdot 
 10^{19}{\rm cm^{-2}}$.
 
 The low SXRB source intensity towards the galactic anticenter can be used 
 to separate the contribution from galactic halo emission and that from 
 unresolved extragalactic point sources. Barber et al. (1996) determined 
 $I_{\rm extragal}\,=\,2.8\cdot10^{-4}\,{\rm cts\,s^{-1}\,arcmin^{-2}}$, 
 while Cui et al. (1996) derived $I_{\rm extragal}\,=\,4.4\cdot10^{-4}\,{\rm 
 cts\,s^{-1}\,arcmin^{-2}}$. Our minimum $\frac{1}{4}$\,keV count rate is 
 about $I_{\rm distant}\,=\,(13\,\pm\,4)\cdot10^{-4}\,{\rm 
 cts\,s^{-1}\,arcmin^{-2}}$. In the extreme cases $I_{\rm distant}^{\rm 
 min}\,=\,9\cdot10^{-4}\,{\rm cts\,s^{-1}\,arcmin^{-2}}$ and $I_{\rm 
 extragal}^{\rm max}\,=\,4.4\cdot10^{-4}\,{\rm cts\,s^{-1}\,arcmin^{-2}}$, 
 the extragalactic X-ray background contribution is about equal to the soft 
 X-ray intensity of the galactic halo. We plotted the $I_{\rm distant}$ 
 values as a function of angular distance from the inner Galaxy in 
 Fig.\,\ref{fig:sxrb_long}. The horizontal lines in the lower part of 
 Fig.\,\ref{fig:sxrb_long} indicate the extragalactic background level 
 determined by Barber et al. (1996) and Cui et al. (1996). $I_{\rm distant}$ 
 increases towards the galactic center. This leads us to conclude that the 
 bulk of the distant soft X-ray emission is of galactic origin and that the 
 extragalactic background radiation gives only a constant X-ray intensity 
 offset.

 \subsection{X-ray enhancements near HVCs}
 
 We have shown that the general radiation transfer of the SXRB photons is 
 well represented by our modelling of the diffuse X-ray background. Our 
 analysis of the {\it ROSAT\/} all-sky data reveals {\em no} evidence for 
 soft X-ray shadows attributable to HVCs.
 This is in some respect surprising because
 our analysis is biased {\em towards the detection} of
 HVC soft X-ray {\em shadows} and, consequently, {\em against} the 
 detection of soft X-ray {\em enhancements} of HVCs (see Sect.\,3.5.3). 
 Certainly, HVCs attenuate the extragalactic background radiation. 
 As shown above, the maximum extragalactic background intensity is 
 $I_{\rm extragal}$\,=\,4.4$\cdot 10^{-4}\,{\rm cts\,s^{-1}arcmin^{-2}}$ 
 (Cui et al. 1996) and an HVC with $N_{\rm HI}({\rm HVC})\,=\,1 \cdot 
 10^{20}\,{\rm cm^{-2}}$ 
 attenuates this radiation by about 60\%. If some HVCs are 
 located at large distances from the galactic disk (see Blitz et al. 1998) 
 this only HVC absorbed X-ray radiation is additionally attenuated by the diffuse galactic \hi 
 layer. This layer may be characterized by a typical $N_{\rm HI}$ of about 
 $N_{\rm HI}({\rm HVC})\,=\,1 \cdot 10^{20}{\rm cm^{-2}}$. 
 {\em On} the HVC we observe a count rate of about $I_{\rm extragal}{(\rm 
 ON)}\,=\,0.9\cdot 10^{-4}\,{\rm cts\,s^{-1}arcmin^{-2}}$ 
 whereas {\em off} the HVC the count rate is $I_{\rm extragal}{(\rm OFF)}\,=\,1.7\cdot 
 10^{-4}\,{\rm cts\,s^{-1}arcmin^{-2}}$. 
 The difference $I_{\rm extragal}{\rm (OFF\,-\,ON)}\,=\,0.8\cdot 
 10^{-4}\,{\rm cts\,s^{-1}arcmin^{-2}}$ is undetectable 
 in the {\it ROSAT\/} data analyzed at the current angular resolution.
 
 \begin{table*}
 \caption[]{Properties of the soft X-ray enhancements towards HVC complexes C, D
 and GCN. The $l$ and $b$ extent is determined by the distribution of
 the $4\sigma$ and $5\sigma$ contour lines, plotted in the individual maps of
 the fields of interest.
 $E_{\rm det}$ denotes the total energy detected by the {\it ROSAT\/} PSPC 
 integrated across the extent of the excess soft X-ray emitting area.
 $\overline{\sigma}$ gives the mean significance level, while
 $\sigma_{\rm max}$ gives the maximum significance level within
 the extent of the excess emission area. 
 To evaluate the emission measure ($EM$) as well as the electron volume
 density ($n_{\rm e}$), we assumed ${\rm log}(T{\rm [K]})\,=\,6.2$ (Kerp et al.
 1998) and a ``normalized'' distance of D\,=\,1\,kpc to the HVCs. For a 
 different distance of the HVCs the radiated 1/4\,keV
 energy ($E_{\rm rad}(1/4{\rm keV}$)) has to be scaled by $D^2$\,[kpc]
 and the corresponding electron density by $D^{-0.5}$\,kpc. The last two
 columns give the field-averaged significance level of the $\frac{1}{4}$\,keV 
 excess emission and the peak significance level.}
 \begin{flushleft}
 \begin{tabular}{lccrrcccr}
 \hline
 complex & {\it l}--range   & {\it b}--range & $E_{\rm det}(1/4{\rm keV})$ & $E_{\rm rad}(1/4{\rm keV})$ & $EM$ & $n_{\rm e}$ & $\overline{\sigma}$ & ${\rm \sigma_{max}}$\\
 & & & ($ 10^{-11}{\rm erg\,cm^{-2}\,s^{-1}}$) & ($10^{34} {\rm erg\, s^{-1}})$ 
 & (${\rm cm^{-6}\,pc}$) & (${\rm cm^{-3}}$) &\\
 \hline
 C & 143$\degr$ -- 148$\degr$ & 42$\degr$ -- 45$\degr$ & 5.4 & 5.0 & 0.054 & 0.03 & 5.0 & 6.3 \\
 C & 129$\degr$ -- 135$\degr$ & 41$\degr$ -- 44$\degr$ & 6.2 & 4.2 & 0.041 & 0.03 & 5.8 & 7.8 \\
 C & 119$\degr$ -- 130$\degr$ & 46$\degr$ -- 51$\degr$ & 21.7 & 6.7 & 0.022 & 0.01 & 6.3 & 8.5 \\
 C & 116$\degr$ -- 122$\degr$ & 39$\degr$ -- 43$\degr$ & 7.3 & 3.9 & 0.034 & 0.02 & 5.6 & 8.2 \\
 C & 110$\degr$ -- 116$\degr$ & 41$\degr$ -- 46$\degr$ & 11.6 & 4.5 & 0.032 & 0.02 & 7.1 & 10.2\\
 C & 110$\degr$ -- 117$\degr$ & 46$\degr$ -- 53$\degr$ & 23.4 & 8.1 & 0.033 & 0.02 & 7.7 & 11.2 \\
 C & 110$\degr$ -- 116$\degr$ & 54$\degr$ -- 58$\degr$ & 7.5 & 2.4 & 0.022 & 0.02 & 5.5 & 7.8 \\
 C & 97$\degr$ -- 111$\degr$ & 50$\degr$ -- 53$\degr$ &  28.5 &  7.6 & 0.031 & 0.02 & 6.5 & 9.0 \\
 C & 99$\degr$ -- 105$\degr$ & 35$\degr$ -- 39$\degr$ &  5.2 & 6.7 & 0.044 & 0.02 & 6.6 & 11.0 \\
 C & 89$\degr$ -- 96$\degr$ & 41$\degr$ -- 45$\degr$ &   14.7 &  3.9 & 0.024 & 0.02 & 6.7 & 10.6 \\
 D & 80$\degr$ -- 84$\degr$ & 23$\degr$ -- 27$\degr$ &   3.7 & 2.4 & 0.028 & 0.02 & 6.6 & 10.1 \\
 GCN & 34$\degr$ -- 40$\degr$ & $-$31$\degr$ -- $-$28$\degr$ & 4.7 & 5.4 & 0.034 & 0.02 & 2.5 & 5.3 \\
 \hline
 \end{tabular}
 \end{flushleft}
 
 \end{table*}
 For the HVC complexes C, A, GCN, and D we found significant 
 soft X-ray emission close to or towards the HVCs. 
 In case of the higher-$l$ end of HVC complex C, 
 soft X-ray enhancements up to the 11-$\sigma$ level were detected.
 The X-ray enhancements generally follow the orientation of the HVCs, 
 for instance in the case of HVC complex C (Fig.\,\ref{fig:hvcc-high}d), 
 but not always in detail.
 In case of HVC complex C, large parts of the complex are located 
 close to intermediate-velocity gas (Kuntz \& Danly 1996).  Depending on 
the origin of the excess soft X-ray radiation, 
 IVCs may also be X-ray bright. To investigate this, we 
 mosaicked the X-ray and \hi data of the entire HVC complex C.  
Figure\,\ref{fig:mosaic} shows a mosaic, where the excess soft X-ray 
emission is  displayed in color and the $N_{\rm HI}$ distribution of the 
HVCs  ($-450 < v_{\rm LSR} < -100\, {\rm km\,s^{-1}}$) and IVCs ($ -75 < 
v_{\rm  LSR} < -25\, {\rm km\,s^{-1}}$) are superposed as contours.
 Contours of the HVC $N_{\rm HI}$ distribution encompass areas of excess 
 X-ray emission (Fig.\,\ref{fig:mosaic}, {\it top}).
 
 IVCs show a lower degree of correlation with the soft X-ray enhancements 
 (Fig.\,\ref{fig:mosaic}, {\it bottom}) than shown by the HVCs 
(Fig.\,\ref{fig:mosaic}, {\it top}). Their \hi emission maxima coincide
 positionally with 
 minima in the X-ray emission, indicating that IVCs absorb the 
 constant $I_{\rm distant}$ intensity distribution. Most probably, they are 
 located nearer than the sources of the excess soft X-rays. 
 Thus, HVCs remain as the most probable candidate for the association with 
 the excess soft X-ray emission, while the role of the IVCs remains 
 unclear. Especially the presence of the ``velocity bridges" (Pietz et al. 
 1996) linking some HVCs with the excess X-ray emission with the 
 intermediate-velocity gas deserves further investigation, especially in 
 regard to the Blitz et al. (1998) predictions.
 
 The X-ray enhancements have a larger angular extent 
 than the HVC \hi distribution. The difference in location between the 
 \hi clouds and the soft X-ray emission is not surprising if we assume that 
 the X-ray emitting plasma and the HVCs are spatially close. Under this 
 hypothesis, the neutral HVC boundaries are ionized by the radiation from the 
 X-ray plasma. This can cause the apparent positional shift between the 
 neutral gas and the X-ray radiation. In consequence, ${\rm H^+}$ radiation 
 should be detectable from this interface region; in particular, it has to 
 originate close to $N_{\rm HI}$ gradients. Towards complexes M, A, and C, 
 Tufte et al. (1998) detected H$\alpha$ radiation. In these cases the soft 
X-ray enhancements reveal the presence of collisionally ionized gas.
 
 These findings can be interpreted in two general ways. First, an \hi  gradient is caused by the ionizing radiation from the X-ray plasma 
 alone (conductive interfaces). Second, an \hi gradient is caused by the 
 interaction of the HVC with the ambient ISM. 
 
 For HVC complex C (higher-$l$ end, Sect.\,4.1.2), 
 we can estimate the energy budget of the apparent interaction process. 
 Assuming that complex C has a total mass of about 
 $M_{\rm HVC}\,\sim\,10^6$ to $10^7{\rm M_{\odot}}$ and a bulk velocity of 
$v_{\rm 
 HVC}\,\sim\,100\,{\rm km\,s^{-1}}$ (Wakker \& van Woerden 1991), the kinetic 
 energy of the complex is $E_{\rm kin}{\rm 
 (HVC)}\,\sim\,10^{53}$ -- $10^{54}$\,erg.  In the {\it ROSAT\/} 
$\frac{1}{4}$\,keV band we detect at maximum 
 $E_{\rm X}\,\sim\,10^{36}\,{\rm erg\,s^{-1}}$. 
 Thus, the observed X-rays require only a very small fraction of the 
 available kinetic energy. 
 This implies that we need, from the energy point of view, 
 only a weakly-efficient process which converts the HVC bulk motion 
 into thermal energy.
 If the excess soft X-ray emission is caused by heating of the HVC and the 
 surrounding medium, we have to investigate the physical conditions of the 
 interaction scenario.
 At a vertical distance of $|z| \sim 3$\,kpc, the temperatures are about 
 $T\,\simeq\,10^5$-$10^6$\,K, the volume densities 
 $n_{\hi}\,\simeq\,10^{-3}{\rm cm^{-3}}$ (Kalberla \& Kerp 1998), and the 
 sound speed $v_{\rm s}\,\simeq\,40$-$120\,{\rm km\,s^{-1}}$. It is 
 difficult to account for a strong shock if the absolute value of the 
 complete HVC velocity vector is $|v_{\rm HVC}|\,\sim\,100\,{\rm 
 km\,s^{-1}}$.
 Two other possibilities are open to overcome this distance discrepancy, 
 namely the galactic wind scenario (Kahn 1991), in which a wind encounters 
 the HVCs, and the magnetic reconnection process (Kahn \& Brett 1993, Zimmer 
 et al. 1997), in which turbulent motions within and close to the HVC disturb 
 the magnetic lines of force. The field lines find a new configuration of 
 minimum energy during the reconnection process. As Zimmer et al. (1997) 
 pointed out, the magnetic reconnection can heat the ISM to several million 
 degrees.
 
 A remaining problem concerns the large angular extent of the areas of 
 excess soft X-ray emission. 
 As Fig.\,\ref{fig:hvcc-high}d shows, the angular extent of the 
 soft X-ray excess emission and the extent of HVC complex C are about equal. 
 Assuming a distance of at least 2\,kpc for complex C, this corresponds 
 to a linear size of at least 150\,pc. 
 Heating such a volume via atomic collisions would require 
 $t_{\rm collision}\,\sim\,5\cdot10^6$\,years.
 This time is comparable to the cooling time of the detected X-ray plasma.
 This may indicate that the thermal expansion of hot gas heated by a single
event may not be the source of the observed X-ray radiation.
 We note that the Alfv\'{e}n 
 velocity is much higher than the sound speed: ${v^2}_{\rm 
 A}\,=\,\frac{B^2}{4\,\pi\,n_{\hi}\,m_{\rm P}} 
 \gg v_{\rm s}$, where $B$ denotes the magnetic field strength and 
 $n_{\hi}$ the volume density of the medium. 
 If $B\,=\,3\,{\rm \mu G}$ (Beuermann et al. 1985) and 
 $n_{\hi}\,=\,1\cdot 10^{-3}{\rm cm^{-3}}$, then 
 ${v}_{\rm A}\,\simeq\,200\,{\rm km\,s^{-1}}$. 
 This indicates that the magnetic lines of force transfer information about 
 the motion of the HVCs in the halo some five times more rapidly than the  particle collisions do. Understanding the role played by 
 magnetic fields may be important to understanding the HVC excess X-ray
 emission scenario.
 
 \section{Summary}
 We compared selected fields from the $\frac{1}{4}$\,keV {\it ROSAT\/} all-sky survey against the 
 Leiden/Dwingeloo \hi survey looking for correlations between HVCs and 
 soft X-ray emission. We considered the soft X-ray radiation transfer in 
 detail towards several prominent HVC complexes. Our results show that:
 
 \begin{enumerate}
 \item The observed SXRB shows a smooth diffuse X-ray source intensity
 distribution at the 
 higher latitudes (see Fig.\,\ref{fig:sxrb_long}).
 \item Small- and large-scale ($0\fdg8$ - $30\degr$) variations of the 
 observed SXRB distribution can be attributed to photoelectric absorption.
 \item \hi alone traces the amount of soft X-ray absorbing interstellar 
 matter well.
 \item Warm \hi (i.e. the intercloud medium) traces most 
 of the soft X-ray absorbing interstellar medium.
 \item The intensity of the distant X-ray emission decreases with increasing 
 angular distance to the galactic center, implying that most of the distant
 soft X-ray emission is galactic in origin. The distant SXRB 
 shows comparable intensities on the northern and southern 
 sky.
 \item The distant soft X-ray emission probably consists of a superposition 
 of a galactic X-ray plasma component ($T_{\rm plasma}\,\sim\,10^{6.3}$\,K) 
 and a component of unresolved extragalactic point sources. More than 50\% of 
 the total radiation observed towards the galactic anticenter can be 
 attributed to the galactic halo plasma emission.
 \end{enumerate}
 
 The above results are consistent with those of Pietz et al. (1998a; 1998b).
 
 We detected the following deviations from the smooth distant galactic X-ray 
 background intensity distribution towards several prominent HVC complexes:
 \begin{enumerate}
 \item Large portions of HVC complex C are positionally associated with 
 excess soft X-ray emission.
 \item Towards a part of HVC complex D, we detected enhanced soft X-ray 
 emission, positionally associated with an HVC filament. However, 
 we note that there is also an IVC close to this area of excess 
 emission.
 \item We detected enhanced soft X-ray emission in the direction of 
 Mrk\,509, where Sembach et al. (1995) found highly-ionized 
 gas associated with HVC complex GCN. 
 \item Towards complex WA we found inconclusive evidence for excess 
 X-rays.
 \end{enumerate}
 
 \begin{acknowledgements}
 J. Kerp thanks the Deutsche Forschungsgemeinschaft for support
 under grant No. ME\,745/17-2. J. Pietz thanks the Deutsche Argentur f\"ur 
 Raumfahrtangelegenheiten for support under grant No. 40012.
 The {\it ROSAT\/} project has been supported by the German
 Bundesministerium f\"ur Bildung, Wissenschaft, Forschung und Technologie (BMBF)
 and by the Max-Planck-Gesellschaft. We thank K.S. de Boer and the anonymous 
 referee for comments based on careful readings of the manuscript.
 \end{acknowledgements}
 
 {}
 \begin{figure*}
 \caption[]{{\it ROSAT\/} $\frac{1}{4}$\,keV photoelectric absorption cross 
 section (Morrison \& McCammon 1983) versus $N_{\rm HI}$. The effective 
 cross section ($\sigma_{\rm X}$) depends on the X-ray spectrum.
 The solid line shows $\sigma_{\rm X}({\rm LHB})$ for the LHB plasma with 
 $T_{\rm LHB}\,=\,10^{5.85}$\,K; the dashed line shows $\sigma_{\rm X}({\rm 
 halo})$ for the distant galactic plasma with $T_{\rm halo}\,=\,10^{6.3}$\,K, 
 based on the Raymond \& Smith (1977) plasma code. The dotted line shows 
 $\sigma_{\rm X}({\rm extragal})$ for the extragalactic power-law spectrum 
 of $E^{-1.5}$ (Gendreau et al. 1995).}
 \label{fig:cross}
 \end{figure*}
 \begin{figure*}
 \caption[]{Maps of the part of HVC complex C at both lower $l$ and lower 
 $b$ (see Sect.\, 4.1.1). {\bf (a)} Observed $\frac{1}{4}$\,keV {\it ROSAT \/}
 PSPC map. Dark colours denote low brightnesses ($I_{\rm 1/4\,kev}({\rm 
 min})\,=\,3.5\cdot 10^{-4}\,{\rm cts\,s^{-1} arcmin^{-2}}$); bright 
 colours, strong emission ($I_{\rm 1/4\,kev}({\rm max})\,=\,17.5 \cdot 
 10^{-4}\,{\rm cts\,s^{-1} arcmin^{-2}}$). Solid lines indicate areas where 
 more soft X-ray emission is observed than expected by the model of 
 Eq.\,(2); dashed lines enclose areas where the X-ray emission is weaker 
 than expected. The lowest contour represents the 4-$\sigma$ level; the 
 contour step is 2 $\sigma$ ($\sigma \approx 0.5 \cdot 10^{-4}\,{\rm 
 cts\,s^{-1}arcmin^{-2}}$). 
 {\bf (b)} Modelled situation which results if a constant SXRB, attenuated only by 
 the $N_{\rm HI}$ values (at $|v_{\rm LSR}| \le 100 {\rm km\,s^{-1}}$) shown 
 in panel (c), with $I_{\rm distant} = (25 \pm 4) \cdot 10^{-4}\,{\rm 
 cts\,s^{-1}arcmin^{-2}}$, is combined with the unabsorbed local radiation 
 of $I_{\rm LHB} = (2.8 \pm 0.5) \cdot 10^{-4}\,{\rm 
 cts\,s^{-1}arcmin^{-2}}$, also assumed constant across the field. The modelled map 
 has the same maximum and minimum intensity values as the {\it ROSAT\/} 
 image shown in panel (a).
 All images have the same angular resolution of 48$\arcmin$.
 {\bf (c)} $N_{\rm HI}$ distribution contributed from $|v_{\rm LSR}| \le 100\, 
 {\rm km\,s^{-1}}$: $N_{\rm HI}({\rm min})\,=\,6\cdot 10^{19}{\rm cm^{-2}}$, 
 $N_{\rm HI}({\rm max})\,=\,6\cdot 10^{20}{\rm cm^{-2}}$.
 {\bf (d)} Greyscale: $N_{\rm HI}$ distribution contributed by HVC velocities, ($-450\,{\rm 
 km\,s^{-1}}\,\leq\,v_{\rm LSR}\,\leq\,-100\,{\rm km\,s^{-1}}$): $N_{\rm 
 HI}({\rm min})\,=\,3\cdot 10^{19}{\rm cm^{-2}}$, $N_{\rm HI}({\rm 
 max})\,=\,6\cdot 10^{20}{\rm cm^{-2}}$. The contours are as described in (a). 
 {\bf (e)} Greyscale: $N_{\rm HI}$ distribution of the IVC velocity regime, ($-75\,{\rm 
 km\,s^{-1}}\,\leq\,v_{\rm LSR}\,\leq\,-25\,{\rm km\,s^{-1}}$): $N_{\rm 
 HI}({\rm min})\,=\,1.5\cdot 10^{19}{\rm cm^{-2}}$, $N_{\rm HI}({\rm 
 max})\,=\,6\cdot 10^{19}{\rm cm^{-2}}$. The contours are as described in (a).
 {\bf (f)} Positional dependence (with galactic longitude [top] and latitude [bottom]) of the averaged modelled and observed SXRB 
 intensity profiles for the lower-longitude end of HVC complex C.
 The solid lines represent the simulated soft X-ray intensity distribution, 
 modelled as described in the text; the points mark the observed 
 distribution and its corresponding 1-$\sigma$ uncertainties.The agreement 
 of the modelled values with those observed indicates that the soft X-ray 
 background radiation is smoothly distributed across the field, and that \hi 
 traces predominately the large-scale photoelectric absorption by the 
 galactic interstellar medium.}
 \label{fig:hvcc-low}
 \end{figure*}
 \begin{figure*}
 \caption[]{
 To constrain the velocity-integration range for $N_{\rm \hi}$,
 $I_{\rm distant}$ is plotted as a function of $N_{\rm \hi}$ for the HVC 
 complex C {\it ROSAT\/} data presented in Fig.\,\ref{fig:hvcc-low}
 (see Sect.\,4.1.2).
 The histograms represent $I_{\rm distant}$ versus $N_{\rm \hi}$ integrated
 over three different velocity ranges. The horizontal solid line marks the
 best-fit $I_{\rm distant}$ intensity level, while the horizontal dashed lines
 indicate the uncertainty range of the modelled value. 
 Within the uncertainties (the errorbar in the lower-right part of the figure), $I_{\rm distant}$ can be
 considered as constant across the field of interest. This finding has two
 major implications; first, $I_{\rm LHB}$ is within the uncertainties also
 constant across the field of interest. Second, the WNM, already enclosed in
 the velocity brackets $|v_{\rm LSR}|\,\leq\,25\,{\rm km\,s^{-1}}$, determines
 the $I_{\rm distant}$ intensity level. The more extreme velocity ranges
 introduce only a minor intensity variation, while the functional dependence
 of $I_{\rm distant}$ on $N_{\rm \hi}$ is unaffected.}
 \label{fig:idist_nh}
 \end{figure*}
 \begin{figure*}
 \caption[]{
 Soft X-ray background towards the higher-$l$ and -$b$ end of HVC complex C, 
 (see Sect. 4.1.2 and 4.1.3). {\bf (a)} The SXRB intensities observed in the {\it 
 ROSAT\/} $\frac{1}{4}$\,keV band. {\bf (b)} The SXRB map modelled according to 
 Eq.\,(2) using Leiden/Dwingeloo data and assuming 
 $I_{\rm distant} = {\rm const.} = (16 \pm 3) \cdot 10^{-4}\, {\rm 
 cts\,s^{-1}arcmin^{-2}}$ in addition to the local X-ray radiation of 
 $I_{\rm LHB}=(3.5\pm0.5)\cdot10^{-4}\,{\rm cts\,s^{-1}arcmin^{-2}}$, 
 also assumed constant across the field. Dark colours denote low X-ray intensities 
 ($I_{\rm 1/4\,keV}({\rm min})\,=\,3.0\cdot 10^{-4} 
 {\rm cts\,s^{-1}arcmin^{-2}}$); bright colours denote high intensities 
 ($I_{\rm 1/4\,keV}({\rm max})\,=\,14.0\cdot 10^{-4} 
 {\rm cts\,s^{-1}arcmin^{-2}}$).
 {\bf (c)} The $N_{\rm HI}$ distribution ($9\cdot 10^{19}\,{\rm 
 cm^{-2}}\,\leq\,N_{\rm HI}\,\leq\,9\cdot 10^{20}\,{\rm cm^{-2}}$) across 
 the field within the range $|v_{\rm LSR}| \leq 100\,{\rm km\,s^{-1}}$.
 {\bf (d)} 
 Greyscale: the $N_{\rm HI}$ distribution ($1\cdot 10^{19}\,{\rm cm^{-2}}\,\leq\,N_{\rm 
 HI}\,\leq\,1\cdot 10^{20}\,{\rm cm^{-2}}$) in the HVC regime, $-450\,{\rm 
 km\,s^{-1}}\,\leq\,v_{\rm LSR}\,\leq\,-100\,{\rm km\,s^{-1}}$. The contours
are described in (a).
 {\bf (e)} Greyscale: the $N_{\rm HI}$ distribution ($7\cdot 10^{19}\,{\rm 
 cm^{-2}}\,\leq\,N_{\rm HI}\,\leq\,2.5\cdot 10^{20}\,{\rm cm^{-2}}$) in the 
 IVC regime, $-75\,{\rm km\,s^{-1}}\,\leq\,v_{\rm LSR}\,\leq\,-25\,{\rm 
 km\,s^{-1}}$. The contours are described in (a).
 {\bf (f)} The intensity profiles averaged in $l$ and $b$ from the maps in panel (a), 
 dots with error-bars, and (b), solid lines. Superposed as contours are the 
 intensity deviations between the observed (a) and modelled (b) SXRB maps. 
 The contours proceed from the 5-$\sigma$ level in steps of 2$\sigma$ 
 ($\sigma \approx 0.65 \cdot 10^{-4} {\rm cts\,s^{-1}arcmin^{-2}}$). 
 Solid contours in (a), (d), and (e) mark areas of excess 
 X-ray emission; dashed contours mark areas of weaker X-ray emission 
 than expected from the map in (b). The angular resolution of the images is 
 48$\arcmin$.
}
 \label{fig:hvcc-high}
 \end{figure*}
 \begin{figure*}
 \caption[]{Maps of the X-ray and \hi sky towards HVC complex GCN (see 
 Sect.\,4.2).
 {\bf (a)} {\it ROSAT\/} $\frac{1}{4}$\,keV SXRB distribution 
 ($I_{\rm 1/4\,keV} ({\rm min})\,=\,2.5\cdot 10^{-4}\,{\rm cts\,s^{-1} 
 arcmin^{-2}}$ and $I_{\rm 1/4\,keV}({\rm max})\,=\,8\cdot 10^{-4}\,{\rm 
 cts\,s^{-1}arcmin^{-2}}$).
 {\bf (b)} Modelled SXRB image derived from the \hi data, assuming both a constant 
 distant X-ray background $I_{\rm distant} = (30\pm 7) \cdot 10^{-4}\,{\rm 
 cts\,s^{-1}arcmin^{-2}}$ 
 and a constant local X-ray source 
 $I_{\rm LHB} = (2.3\pm 0.5) \cdot 10^{-4}\,{\rm cts\,s^{-1}arcmin^{-2}}$.
 Solid contours indicate excess X-ray emission; dashed contours indicate an
 emission deficiency. The contours proceed from the 4-$\sigma$ level in 2-$\sigma$ 
 steps ($\sigma \approx 0.8 \cdot 10^{-4}\,{\rm cts\,s^{-1}arcmin^{-2}}$).
 {\bf (c)} $N_{\rm HI}$ distribution of the soft X-ray absorbing ISM ($|v_{\rm 
 LSR}|\,\leq\,100\,{\rm km\,s^{-1}}$ colour coded within the range $3\cdot 
 10^{20}\,{\rm cm^{-2}}\,\leq\,N_{\rm HI}\,\leq\,1\cdot 10^{21}\,{\rm 
 cm^{-2}}$.
 {\bf (d)} Greyscale: the HVC $N_{\rm HI}$ distribution ($-450\,{\rm km\,s^{-1}}\,\leq\,v_{\rm 
 LSR}\,\leq\,-100\,{\rm km\,s^{-1}}$ and $4\cdot 10^{18}\,{\rm 
 cm^{-2}}\,\leq\,N_{\rm HI}\,\leq\,1\cdot 10^{19}\,{\rm cm^{-2}}$). The contours are described in (a).
 {\bf (e)} IVC $N_{\rm HI}$ distribution 
 ($25\,{\rm km\,s^{-1}}\,\leq\,|v_{\rm LSR}|\,\leq\,75\,{\rm km\,s^{-1}}$ 
 and $2.5\cdot 10^{19}\,{\rm cm^{-2}}\,\leq\,N_{\rm HI}\,\leq\,7.5\cdot 
 10^{19}\,{\rm cm^{-2}}$). The contours are described in (a).
 {\bf (f)} SXRB intensity profiles, averaged in $l$ and $b$ across the map (panel 
 (a): dots and error bars; panel (b): solid lines).
 The images in (a) and (b) are scaled identically and have an angular 
 resolution of 48$\arcmin$.
 The dot in (a) marks the position of Mrk\,509 where Sembach et al. (1995) 
 detected highly-ionized high-velocity gas in HST absorption-line 
 measurements, at a location coinciding with excess soft X-ray emission.
 }
 \label{fig:gcn}
 \end{figure*}
 \begin{figure*}
 \caption[]{Maps of the X-ray and \hi sky towards HVC complex WA (see 
 Sect.\,4.3).
 {\bf (a)} Observed $\frac{1}{4}$\,keV SXRB distribution 
 ($I_{\rm 1/4\,keV} ({\rm min})\,=\,4.3\cdot 10^{-4}{\rm 
 cts\,s^{-1}arcmin^{-2}}$ and $I_{\rm 1/4\,keV}({\rm max})\,=\,10\cdot 
 10^{-4}\,{\rm cts\,s^{-1}arcmin^{-2}}$).
 {\bf (b)} Modelled SXRB image derived from the \hi data, assuming a constant 
 intensity distribution across the field of both X-ray source terms, $I_{\rm 
 distant}\,=\,(13\,\pm\,4)\cdot 10^{-4}\,{\rm cts\,s^{-1}arcmin^{-2}}$
 and $I_{\rm LHB}\,=\,(4.3\,\pm\,0.5)\cdot 10^{-4}\,{\rm 
 cts\,s^{-1}arcmin^{-2}}$. Images (a) and (b) are scaled identically; the 
 angular resolution of the maps is 48$\arcmin$. Solid contours indicate 
 excess X-ray emission; dashed contours, a lack of emission. The contours 
 proceed from the 4-$\sigma$ level in steps of 2$\sigma$, where $\sigma 
 \approx 0.8 \cdot 10^{-4}\,{\rm cts\,s^{-1}arcmin^{-2}}$.
 {\bf (c)} Distribution of $N_{\rm HI}$ in the range $|v_{\rm LSR}| \leq 100\,{\rm 
 km\,s^{-1}}$ with $1.5\cdot 10^{20}\,{\rm cm^{-2}}\,\leq\,N_{\rm 
 HI}\,\leq\,8\cdot 10^{20}\,{\rm cm^{-2}}$.
 {\bf (d)} Greyscale: the distribution of $N_{\rm HI}$ in the appropriate positive-velocity HVC 
 range ($100\,{\rm km\,s^{-1}}\,\leq\,v_{\rm LSR}\,\leq\,400\,{\rm 
 km\,s^{-1}}$ with $5\cdot 10^{18}\,{\rm cm^{-2}}\,\leq\,N_{\rm 
 HI}\,\leq\,3\cdot 10^{19}\,{\rm cm^{-2}}$). The contours are described in (a).
 {\bf (e)} Greyscale: the distribution of $N_{\rm HI}$ in the IVC range ($25\,{\rm 
 km\,s^{-1}}\,\leq\,v_{\rm LSR}\,\leq\,75\,{\rm km\,s^{-1}}$ with $5\cdot 
 10^{18}\,{\rm cm^{-2}}\,\leq\,N_{\rm HI}\,\leq\,1.7\cdot 10^{20}\,{\rm 
 cm^{-2}}$). The contours are described in (a).
 {\bf (f)} SXRB intensity averaged over $l$ and $b$. The dots and error bars refer 
 to the observed map in (a); the solid line represents the model in (b).}
 \label{fig:WA}
 \end{figure*}
 \begin{figure*}
 \caption[]{Dependence of $I_{\rm distant}$ (dots) derived from 
 our analysis on angular distance from the galactic center. The data point 
 without an error bar corresponds to an averaged value extracted from the 
 analysis of Herbstmeier et al. (1995). With the exception of this point, 
 all $I_{\rm halo}$ intensities are towards comparable $b$. The horizontal 
 solid and dashed lines represent the $I_{\rm extragal}$ intensity level 
 based on Barber et al. (1996) and Cui et al. (1996), respectively. $I_{\rm 
 distant}$ shows a continuous decrease with increasing angular distance from
 the galactic center and is significantly larger than $I_{\rm 
 extragal}$ towards all analyzed fields.}
 \label{fig:sxrb_long}
 \end{figure*}
 \begin{figure*}
 \caption[]{Mosaic showing the positional correlation of excess 
 $\frac{1}{4}$\,keV emission and the HVC and IVC $N_{\rm HI}$ distributions 
 towards the entire HVC complex C. The images present the areas of excess soft X-ray emission in the significance range 4$\sigma$ (dark color) to 10$\sigma$ (bright color).  {\bf Top:} The HVC $N_{\rm HI}$ distribution 
 ($ -450 < v_{\rm LSR} < -100\, {\rm km\,s^{-1}}$) superposed as contours 
 with $1\cdot 10^{19}\,{\rm cm^{-2}}\,\leq\, N_{\rm HI}\,\leq\,1\cdot 
 10^{20}\,{\rm cm^{-2}}$ in steps of $\Delta N_{\rm HI}\,=\,1\cdot 
 10^{19}\,{\rm cm^{-2}}$.
 {\bf Bottom:} The IVC $N_{\rm HI}$ distribution ($-75 < v_{\rm LSR} < -25\, 
 {\rm km\,s^{-1}}$) superposed as contours with $5\cdot 10^{19}\,{\rm 
 cm^{-2}}\,\leq\, N_{\rm HI}\,\leq\,2\cdot 10^{20}\,{\rm cm^{-2}}$ in steps 
 of $\Delta N_{\rm HI}\,=\,2.5\cdot 10^{19}\,{\rm cm^{-2}}$.
 The HVC $N_{\rm HI}$ distribution follows the orientation of the soft X-ray 
 enhancements. The IVCs reach $N_{\rm HI}$ maxima not positionally coincident with 
 excess X-ray emitting areas, except near $l\,\sim\,102\degr$, $b\,\sim\,37
 \degr$, close to the Draco nebula, and near $l\,\sim\,118\degr$, $b\,
\sim\,42\degr$. Both IVCs are located close to HVCs. In particular, the cloud 
at $l\,\sim\,102\degr$, $b\,\sim\,37\degr$ is close to an \hi velocity bridge 
(VB\,111+35, Pietz  et al. 1996) which was not detectable in the higher $l$ 
and $b$ portion  of HVC complex C (Sect.\,4.1.2). A further investigation of 
the connection of IVCs and HVCs is mandatory.}
 \label{fig:mosaic}
 \end{figure*}
\end{document}